\RequirePackage{rotating}

\documentclass[fleqn,usenatbib]{mnras}

\usepackage{newtxtext,newtxmath}

\usepackage[T1]{fontenc}

\DeclareRobustCommand{\VAN}[3]{#2}
\let\VANthebibliography\thebibliography
\def\thebibliography{\DeclareRobustCommand{\VAN}[3]{##3}\VANthebibliography}

\usepackage{graphicx}	
\usepackage{amsmath}	
\usepackage{amssymb}	
\usepackage{comment}
\usepackage{todonotes}
\usepackage{longtable}
\usepackage{threeparttable}
\usepackage{lastpage}
\usepackage{pbox}
\usepackage{pdflscape}
\usepackage{caption}
\newcommand{\uJy}{$\mu$Jy}
\newcommand{\uJybeam}{$\mu$Jy\,beam$^{-1}$}
\newcommand{\mJybeam}{mJy\,beam$^{-1}$}

\newcommand\Fint{F_{\mathrm{int}}}
\newcommand\Fpeak{F_{\mathrm{peak}}}
\newcommand\pb[2]{\pbox{#1}{\strut #2\strut}}
\newcommand\phn{\phantom{0}}
\newcommand\nodata{...}
\newcommand\NiiL{[\ion{N}{2}] $\lambda\lambda 6548, 6583$}

\def\ha{{H$\alpha$}}
\def\hb{{H$\beta$}}

\makeatletter%
\@ifclassloaded{mnras}%
{%

\def\NiiL{[\ion{N}{ii}] $\lambda\lambda 6548, 6583$}

\def\sii{[\ion{S}{ii}]}

\def\oiii{[\ion{O}{iii}]}
\def\feii{[\ion{Fe}{ii}]}
\def\hii{\ion{H}{ii}}
}%
{%

\def\NiiL{[\ion{N}{2}] $\lambda\lambda 6548, 6583$}

\def\sii{[\ion{S}{2}]}

\def\oiii{[\ion{O}{3}]}
\def\feii{[\ion{Fe}{2}]}
\def\hii{\ion{H}{2}}
}%
\makeatother%

\newcommand{\MSOL}{\mbox{$\:M_{\sun}$}}

\newcommand{\EXPU}[3]{\mbox{\rm $#1 \times 10^{#2} \rm\:#3$}}  

\def\ha{{H$\alpha$}}
\def\hb{{H$\beta$}}

\graphicspath{{./}{figures/}}

\title[Radio catalogue of M83]{A New Radio Catalogue for M83: Supernova Remnants and H~II Regions}

\author[T. D. Russell et al.]{Thomas D. Russell,$^{1}$\thanks{E-mail: t.d.russell@uva.nl}
Richard L. White,$^{2}$
Knox S. Long,$^{2,3}$
William P. Blair,$^{4}$
\newauthor Roberto Soria,$^{5,6}$ and
P. Frank Winkler$^{7}$
\\
$^{1}$Anton Pannekoek Institute for Astronomy, University of Amsterdam, 1098 XH Amsterdam, The Netherlands\\
$^{2}$Space Telescope Science Institute, Baltimore, MD, USA\\
$^{3}$Eureka Scientific, Inc., 2452 Delmer Street, Suite 100, Oakland, CA 94602-3017, USA\\
$^{4}$Department of Physics \& Astronomy, Johns Hopkins University, Baltimore, MD, USA\\
$^{5}$College of Astronomy and Space Sciences, University of the Chinese Academy of Sciences, Beijing 100049, China\\
$^{6}$Sydney Institute for Astronomy, The University of Sydney, Sydney, NSW 2006, Australia\\
$^{7}$Department of Physics, Middlebury College, Middlebury, VT, USA\\
}

\pubyear{2020}

\begin{document}
\label{firstpage}
\pagerange{\pageref{firstpage}--\pageref{LastPage}}
\maketitle

\begin{abstract}
We present a new catalogue of radio sources in the face-on spiral galaxy M83. Radio observations taken in 2011, 2015, and 2017 with the Australia Telescope Compact Array (ATCA) at 5.5 and 9 GHz have detected 270 radio sources.  Although a small number of these sources are background extragalactic sources, most are either \hii\  regions or supernova remnants (SNRs) within M83 itself. Three of the six historical supernovae are detected, as is the very young remnant that had been identified in a recent study, which is likely the result of a supernova that exploded in the last $\sim$100 years but was missed. All of these objects are generally fading with time.  Confusion limits our ability to measure the radio emission from a number of the SNRs in M83, but 64 were detected in unconfused regions, and these have the approximate power-law luminosity function which has been observed in other galaxies. The SNRs in M83 are systematically smaller in diameter and brighter than those that have been detected at radio wavelengths in M33.  A number of the radio sources are coincident with X-ray sources in M83; most of these coincident sources  turn out to be supernova remnants. Our dual frequency observations are among the most sensitive to date for a spiral galaxy outside the Local Group; despite this we were not able to place realistic constraints on the spectral indices, and as a result, it was not possible to search for supernova remnants based on their radio properties alone.

\end{abstract}




\section{Introduction}

Supernova remnants (SNRs) arise from the interaction of matter ejected by a supernova (SN) explosion with the surrounding circumstellar (CSM) and interstellar medium (ISM).  The primary shock from a SNR is primarily observed at radio and X-ray wavelengths, while secondary shocks traversing denser knots of the ISM produce ultraviolet, optical, and infrared line emission.  In the Galaxy, approximately 325 SNRs are known, of which  most  were first identified as radio sources.  Today, however, far more SNRs are known in external galaxies, and most of these were discovered optically through the diagnostic ratio of the optical \sii$\lambda\lambda$6716,6731 lines to \ha\ \cite[see, e.g.][and references therein]{long17}.  Emission from the shock-heated gas characterising SNRs typically has  \sii:\ha\ ratio greater than 0.4, the result of impulsive (shock) heating followed by a long cooling tail behind supernova shocks, whose emission is  characterized by forbidden lines from a range of ionization states, including S$^+$.  In contrast, photoionization by hot stars maintains \hii\ regions in higher ionization states, and the  \sii:\ha\ ratio is typically $\lesssim 0.2$.

For understanding SNR populations and the underlying physics that produce them, samples in nearby galaxies have two important advantages: (1) all of the objects in a single galaxy are at essentially the same distance, and (2) absorption (at least for galaxies with low inclination) is less of an issue than it is for comparing the relative properties of SNRs in the Milky Way.  However, in order to fully characterise extragalactic samples, detections at X-ray and radio wavelengths are essential, and the vast majority of SNRs in galaxies beyond the Local Group have not been detected either at radio wavelengths or in X-rays due primarily to limitations in sensitivity and angular resolution.

M83 (NGC~5236) is a superb laboratory for the study of SNRs; it is a relatively nearby (4.61 Mpc; \citealt{saha06}), nearly face-on, spiral galaxy with a starburst nuclear region and active star formation throughout its prodigious spiral arm structure.  Consistent with its high star formation rate of 3-4 \MSOL ~ yr$^{-1}$  \citep{boissier05}, it has hosted six historically recorded SNe within the past century \citep[e.g.,][]{stockdale06}, probably plus another that was not detected \citep{blair15}.   Only one other galaxy (NGC~6946) is known to have been more prolific in SN production.
As a result of a series of studies using large ground-based telescopes and the {\em Hubble} Space Telescope ({\it HST}), more than 300 SNRs have been identified in M83, more than for any other external galaxy \citep{dopita10, blair12, blair14, winkler17, garofali17, williams19}. Furthermore, through deep X-ray observations of M83 with {\em Chandra},  X-ray counterparts to about 87 of these SNRs have been detected \citep{long14}.  

The first radio observations of M83 were carried out in 1981 using the not-yet-completed Karl G. Jansky Very Large Array (VLA), motivated largely by a search for the radio remnants of M83's historical SNe \citep[five at the time, of which none were then detected,][]{cowan82}.  Subsequent VLA observations  at both 20 cm and 6 cm  were carried out in 1983, 1990, and 1998, all using hybrid array configurations to observe this southern galaxy from the northern VLA, as summarized by \citet{maddox06}.  These later observations detected a total of 55 compact sources, including four of the six historical SNe in M83 \citep[][and references therein]{stockdale06}, plus, as it turns out, the object later identified as a very young SNR by \citet[see below]{blair15}.  Four additional radio sources coincided with SNRs that had been detected optically by \citet{blair04}, of which three also have corresponding X-ray sources found in {\em Chandra} observations by \citet{soria03}.

In \citet{long14}, we presented a set of initial radio observations taken in 2011 with the Australia Telescope Compact Array (ATCA). With those observations, we identified 47 radio sources that were positionally coincident with X-ray or optically-identified SNRs (or SNR candidates) in M83. In this work, we report on a deeper set of radio observations of M83 with the inclusion of ATCA radio observations taken in 2015 and 2017 (in addition to the 2011 results). These observations were primarily intended to determine the radio properties of the SNR population of M83, and secondarily to characterize radio emission in M83 more generally.  We describe the  observations  and the reduction of the data in Section \ref{sec:ATCA}.  We report on our effort  to generate catalogues and extract fluxes for known source populations  from the radio images and our difficulties in determining spectral indices in Section \ref{sec:catalogues}.  We compare our results to those obtained by \citet{maddox06} and discuss our attempts to identify radio sources with known astrophysical objects in  Section \ref{sec:results}.
We explore some of the implications of the identifications for our understanding of the radio properties of SNRs and historical SNe of M83 in Section \ref{sec:discussion},  and finally we summarize what we have learned in Section \ref{sec:summary}.

\section{ATCA Observations and Data Processing}
\label{sec:ATCA}

M83 was observed with the Australia Telescope Compact Array (ATCA) on 2011 April 28, 29, and 30, 2015 June 24, 25, and 26, and 2017 November 20, 21, and 22, all under project code C2494.  For the 2011 and 2015 observations the array was in its most-extended 6\,km (6A and 6D) configurations. Our 2017 ATCA observations were taken with the array in a more compact 1.5\,km (1.5A) configuration to provide improved sensitivity to larger structures.  Table~\ref{tab:radio_obs} gives a journal of these observations and the array configuration used for each. Results from the 2011 observations  have already been presented by \citet{long14}. In this paper, we report cumulative results from the full set of observations at all three epochs, including those from 2011.
All the ATCA observations were centred at 5.5 and 9.0~GHz (recorded simultaneously), with a bandwidth of 2~GHz at each frequency. Observations on each of the nine days were for a duration of 12 hours, providing a total observing time of 108\,hours. 

The data were processed following standard procedures in \texttt{Miriad} \citep{sault95}. We used the ATCA primary calibrator PKS 1934$-$638 for both bandpass and flux density calibration, while the nearby source J1313$-$333 was used for phase calibration. Calibrated data were then imported into the Common Astronomy Software Application (\texttt{CASA}; \citealt{mcmullin07}) for image construction. To account for all of the extended emission, deconvolution was carried out with the multi-scale clean algorithm \citep{rich08} within the \texttt{CASA} task \textsc{CLEAN}, which models the emission as a collection of different scales while preserving extended structure. The images were produced with a Briggs weighting robustness of -1 to balance sensitivity and resolution while suppressing the sidelobes created by the bright nuclear region. 

\begin{table}
\caption{Journal of the ATCA radio observations, showing the on-source observation start time and end times, as well as the telescope configuration. All observations were recorded at central frequencies of 5.5 and 9~GHz.}
\centering
\label{tab:radio_obs}
\begin{tabular}{ccc}
\hline
Start time   & End time & Telescope \\
 (UT)        & (UT)     & configuration$^a$ \\

\hline
2011-04-28 08:12:10 & 2011-04-28 18:35:30 & 6A \\
2011-04-29 07:59:10 & 2011-04-29 17:43:30 & 6A \\
2011-04-30 08:56:20 & 2011-04-30 17:31:20 & 6A \\

2015-06-24 03:49:00 & 2015-06-24 14:57:20 & 6D \\
2015-06-25 03:43:10 & 2015-06-25 14:55:50 & 6D \\
2015-06-26 03:48:00 & 2015-06-26 14:51:10 & 6D \\

2017-11-20 18:24:10 & 2017-11-21 05:22:30 & 1.5A \\
2017-11-21 17:30:40 & 2017-11-22 05:25:20 & 1.5A \\
2017-11-22 17:51:20 & 2017-11-23 05:25:10 & 1.5A \\

\hline
\multicolumn{3}{l}{  
\begin{minipage}{0.9\columnwidth} $^a$ https://www.narrabri.atnf.csiro.au/operations/\\array\_configurations/configurations.html
\end{minipage}
}
\end{tabular}
\end{table}

\begin{figure*}
\centering
\includegraphics[width=1\textwidth]{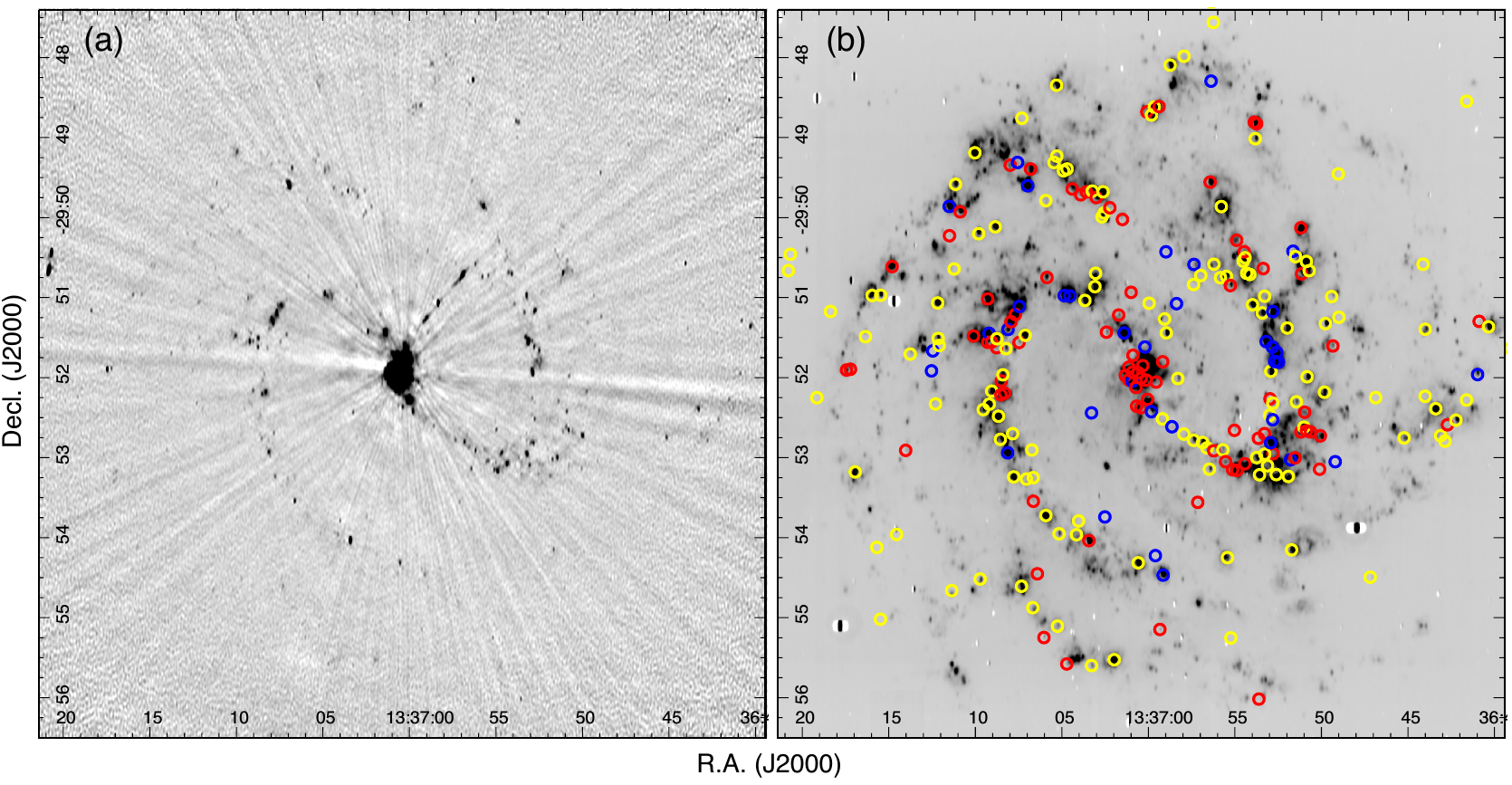}
\caption{An overview comparison of radio and \ha\ emission in M83. (a) the  M83 detection image, where the 5.5 and 9~GHz data have been stacked to provide the most sensitive and highest resolution data set.  (b) A  continuum-subtracted \ha\ image of M83 from the Magellan telescope \citep{blair12}, smoothed to match the size of the radio beam.  Sources from the radio catalogue are marked as follows: sources identified with SNRs are indicated by red circles;  ones that are not SNRs but are associated with X-ray sources are shown in blue; and  the remaining radio sources are shown in yellow.
\label{fig_overview}}
\end{figure*}

\begin{figure}
\centering
\includegraphics[width=0.95\columnwidth]{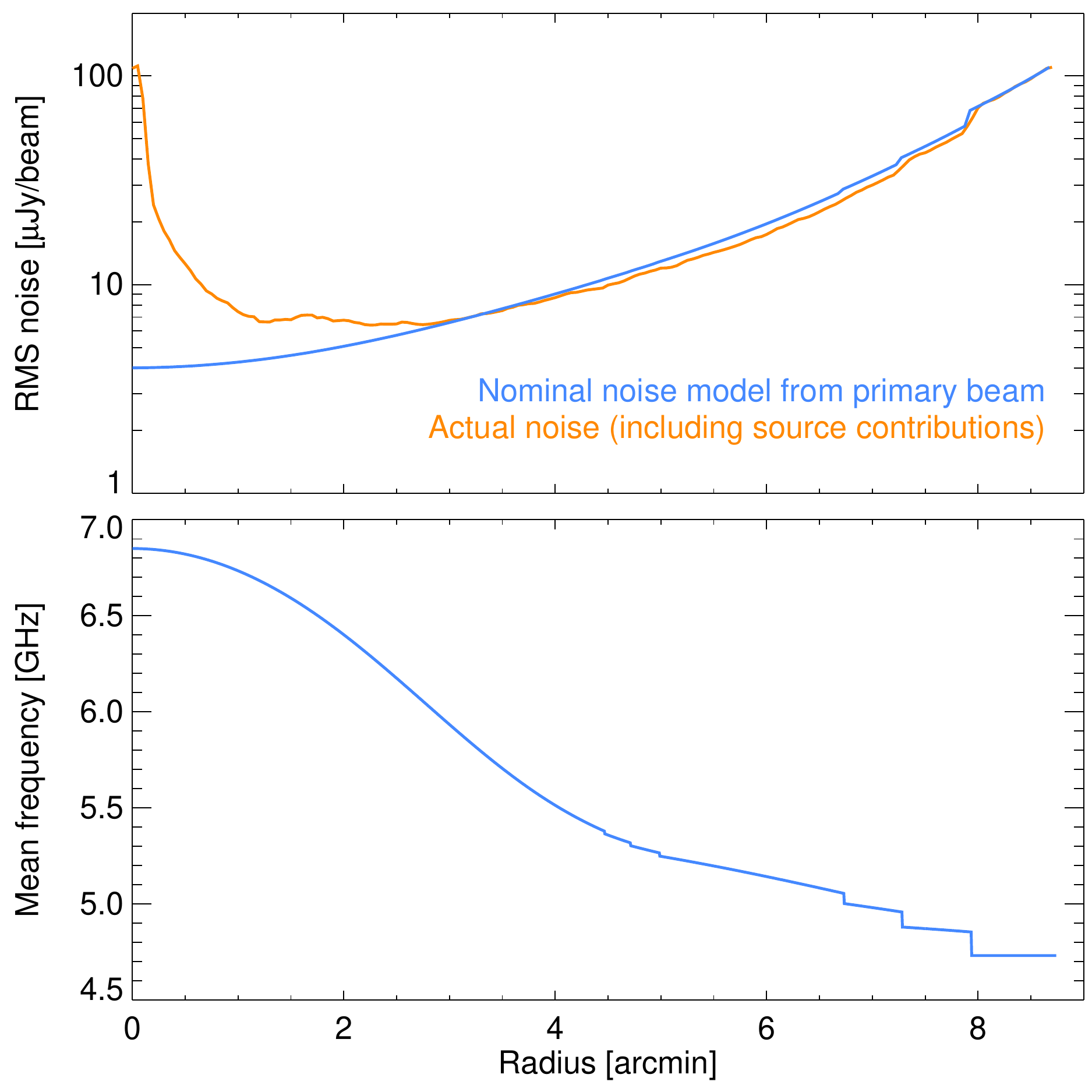}
	\caption{Dependence of rms noise (top) and effective frequency (bottom) in the primary-beam-weighted detection image as a function of field center distance.  
	The primary beam response is narrower at higher frequencies, which both reduces the sensitivity and shifts the mean frequency toward lower values
	at larger off-axis distances.  Small jumps occur at the edges of coverage for the frequency sub-bands.
	The simple model assuming the map noise was uniform before the primary beam correction (blue line) is a good approximation in the outer parts of the galaxy, but near the nucleus the noise contribution from bright sources increases the actual noise in the map sharply (orange line).
\label{fig:noise_vs_dist}}
\end{figure}

Images were processed using four 0.5-GHz sub-bands within each of the two frequency bands, which had elliptical synthesized beams that range from $1.5\times0.7$\arcsec\ (FWHM) at the highest frequency to $3.2\times1.4$\arcsec\ at the lowest frequency. Due to the presence of radio frequency interference in the lowest 0.5-GHz portion of the 9~GHz band, this sub-band was omitted from subsequent analysis; therefore, 7 sub-bands are used in our catalogues (four 0.5-GHz sub-bands centred at 5.5 GHz, and three 0.5-GHz sub-bands centred at 9.2~GHz).\footnote{For convenience, we continue to refer to the latter band as the 9 GHz band below unless this detail is relevant.}  After cleaning, the images were convolved with an elliptical Gaussian to produce a constant resolution of $3.2\times1.4$\arcsec\ (oriented north-south) across the entire bandpass. The final images were primary-beam corrected because of the large angular size of M83 (in comparison to the ATCA primary beam). During tests of the radio spectral index (see Section~\ref{sec:spectralindices} for full discussion), we also applied wavelength-based cuts and tapering to the uv-data to best match the spatial sensitivities between different frequency sub-bands. However, they did not change our final results. Therefore, for our full catalogue we used our most sensitive images, which did not include uv-cuts or tapering.  

\begin{figure*}
	\centering
\includegraphics[width=0.95\textwidth]{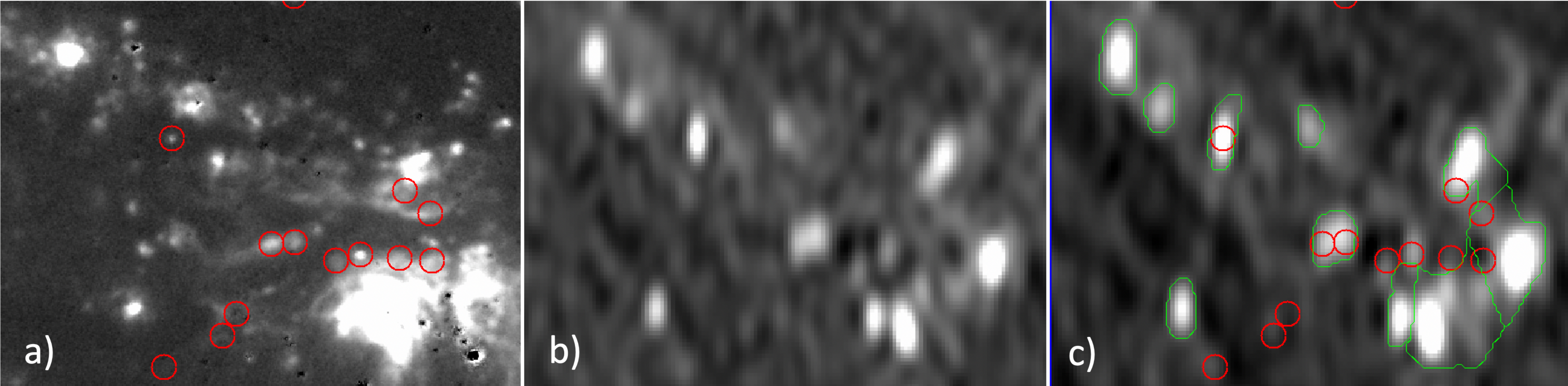}
\caption{Section of M83 SW of the nucleus centered at RA(J2000) 13:36:55.7, Dec(J2000) -29:53:00.1, showing a complicated \hii\ region complex.  The three panels are a) the continuum-subtracted \ha\ data from Magellan \citep{blair12}; b) the ATCA 5.5 GHz data, and c) the combined radio detection image (5.5 GHz + 9 GHz) as described in the text.  The region shown is 30\arcsec\ in the N-S dimension and red circles are 2\arcsec\ in diameter, indicating optical SNR positions.  The green regions in the right panel indicate the radio source islands discussed in the text.
\label{fig_rad_islands}}
\end{figure*}

To provide the maximum sensitivity for source detection, the seven radio sub-band images with matched resolution were combined using optimal weighting, where each sub-band is weighted by the square of the frequency-dependent primary beam response \citep[see][]{becker95}. This combined radio map provides the deepest look, and we refer to it as the detection image in what follows. We experimented with different weighting schemes for the sub-bands, including the possibility of down-weighting or excluding the 9 GHz bandpass altogether in the detection image. Those experiments led us to exclude the low frequency sub-band of the 9 GHz bandpass.  Other changes had little effect on the list of detected sources. Figure~\ref{fig_overview} shows the final detection radio image of M83 (left panel) and a continuum-subtracted  \ha\ image of the galaxy from Magellan/IMACS, smoothed to match the ATCA image resolution \citep[][shown in the right panel]{blair12}. Radio emission is concentrated along the spiral arms of the galaxy, as one would expect if most of the emission arises either from \hii\ regions or SNRs, both of which are associated with regions of star formation visible in the right panel.

The primary beam diameter in the detection image decreases as the inverse frequency, so this radio detection image has an effective frequency that varies with the distance from the field center, from 6.8 GHz at the galaxy center to 4.7 GHz at a radius of 8\arcmin.  The rms noise in this detection image also increases with field-center distance as the primary beam response and available frequency sub-bands drop off (see Figure~\ref{fig:noise_vs_dist}).

To understand the steeper than expected radio spectral indices we derive (see Section~\ref{sec:spectralindices} for full discussion), results were checked for phase decorrelation, which could artificially steepen the observed spectrum of each source. To do this, we re-calibrated the data treating every second scan of the phase calibrator as a target and every other scan as the secondary calibrator. We do not find obvious phase decorrelation from this test.

\section{Source Detection and Flux determinations}

\label{sec:catalogues}

\subsection{Radio Source Catalogue}

We have constructed a catalogue of compact sources in our radio images following the approach used by \citet{white19} to analyze radio observations of M33 with the VLA.  The radio detection image
that combines both the 5.5 and 9 GHz data was processed by a multi-resolution median
pyramid algorithm to separate it into a stack of images that have
structures on scales ranging from compact to very extended.  For each level of the
stack, the local rms noise was
estimated and a segmentation map was constructed, where
contiguous pixels above the noise threshold were grouped into ``islands''
for different sources (see Figure~\ref{fig_rad_islands}).  An iterative approach was used to remove objects detected
in the high-resolution channels of the stack so that they did not contaminate the
lower-resolution channels. 

Flux densities were measured by integrating the multi-resolution
stacked images (which are effectively background subtracted) over the regions defined by
the matched multi-resolution segmentation map.  Separate multi-resolution stacks were
computed for each frequency channel, but the same segmentation map was used for all frequency bands.
The consistently matched image resolutions, median filtering, and segmentation maps
across all frequencies led to consistent, unbiased flux densities with good signal-to-noise and
accurate measurements for sources having irregular morphologies in crowded fields.  See
section 2.1 and the appendix of \citet{white19} for more details on the source finding technique.

Figure~\ref{fig_rad_islands} shows an enlargement of a region in the SW spiral arm, comparing the \ha\ emission in the region to the radio emission.  The red circles in the left and right panels show the positions of optical SNRs for reference, while the middle panel shows the 5.5\,GHz emission without regions, to show the full details without overlays.  The green irregular regions in the right panel show the radio islands described above.  In the crowded region at lower right, one can see how an individual island may incorporate emission from more than a single source that would be picked up by eye.  This is why we have also made catalogues based on forced photometry at the positions of known objects of interest as well (see section~\ref{sec:forced_photometry}).

\begin{figure*}
	\centering
\includegraphics[width=0.95\textwidth]{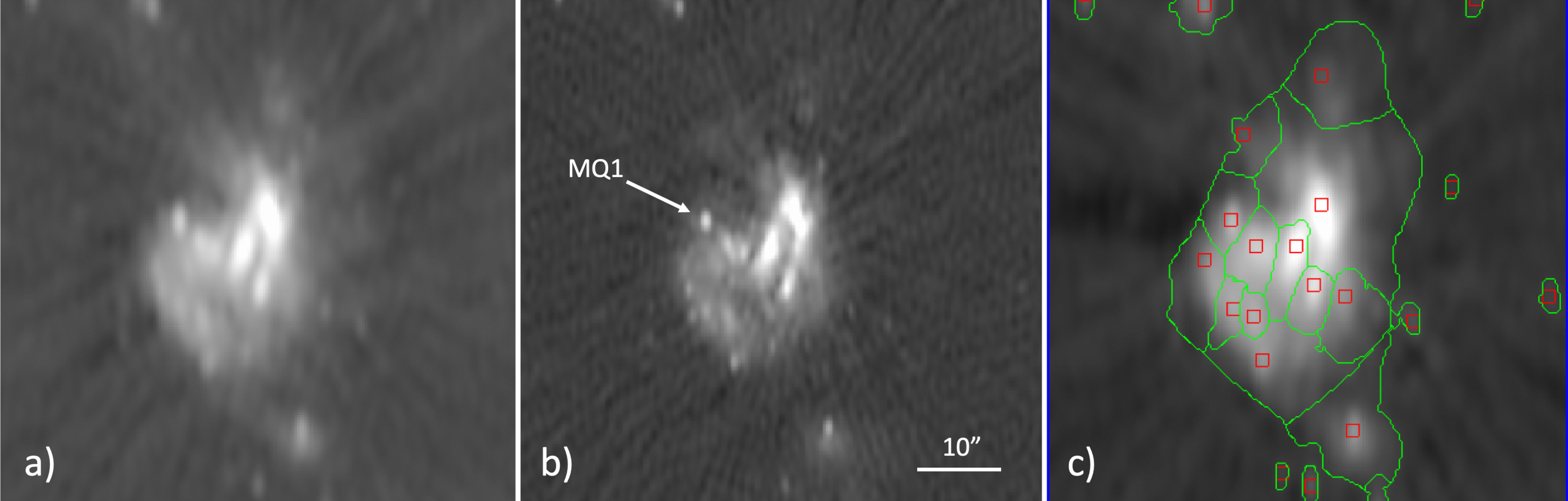}
\caption{This figure shows a 1\arcmin\ region centered on RA(J2000) 13:37:00.7, Dec(J2000) -29:51:54.8, showing the crowded and complex M83 nuclear region.  The three panels are a) the ATCA 5.5 GHz data; b) the ATCA 9 GHz data, and c) the combined radio detection image (5.5 GHz + 9 GHz) as described in the text.   In this case, the islands abut one another and individual sources are not confidently resolved in most cases.  The improved resolution in the 9 GHz data still does not resolve many of the structures in the nuclear region, although the microquasar MQ1 just outside the bright, confused nucleus is indicated in the middle panel.  The red squares are the median positions of each island, as reported in the catalogue.
\label{fig_nuc_islands}}
\end{figure*}

As shown in Figure~\ref{fig_nuc_islands}, the nuclear region of M83 is a bright and complex region in our radio maps, with bright diffuse emission and complex structure making reliable identifications of sources and measurements of the flux densities difficult.  Here the islands actually abut one another due to the to the complexity of the region.  As a result, while we have kept the sources in this region in the catalogue, they are flagged as nuclear sources and should be ignored for most purposes.  We discuss the nuclear region and some of its sources in more detail in Section~\ref{sec:nucleus} below.

\begin{sidewaystable*}
\setlength{\tabcolsep}{0.5pt}
\scriptsize
\begin{center}
\caption{Catalogue of M83 Radio Sources \label{tab:radio_catalogue}}
\begin{tabular}{cccccccccccrccccccccc}
\hline
\hline
Name & RA & Dec & Wrn & F5.5GHz & F9GHz & $\Fint$ & $\nu_p$ & nBands & major & minor & PA & Island & $\Fpeak$ & H$\alpha(\mathrm{tot})$ & H$\alpha(\mathrm{ave})$ & AltName & Sflg & SNRname1 & SNRname2 & nXray \\
~ & (J2000) & (J2000) & ~ & ($\mu$Jy) & ($\mu$Jy) & ($\mu$Jy) & (GHz) & ~ & (arcsec) & (arcsec) & (deg) & ~ & (\uJybeam) & \pbox{5em}{\centering $(\mathrm{erg}\,\mathrm{cm}^{-2}$\break$\mathrm{s}^{-1})$} & \pbox{5em}{\centering $(\mathrm{erg}\,\mathrm{cm}^{-2}$\break$\mathrm{s}^{-1}\mathrm{arcsec}^{-2})$} & ~ & ~ & ~ & ~ & ~ \\
\hline
R20-104 & 13:36:56.252 & -29:47:33.71 & Ec & \phn\phn\phn83.2 $\pm$ \phn11.7 & \phn\phn\phn-3.8 $\pm$ \phn36.7 & \phn\phn\phn89.7 $\pm$ \phn11.8 & 5.277 & 6 & 3.0 & 1.2 & 5.1 & 104 & \phn\phn89.6 $\pm$ \phn9.4 & 3.5e-17 & 5.7e-18 & \nodata & 0 & \nodata & \nodata & 0 \\
R20-105 & 13:36:56.368 & -29:47:18.45 & Ec & \phn\phn\phn90.4 $\pm$ \phn11.4 & \phn\phn\phn80.4 $\pm$ \phn30.2 & \phn\phn\phn90.4 $\pm$ \phn11.4 & 5.516 & 5 & 3.2 & 1.4 & 163.8 & 105 & \phn\phn81.0 $\pm$ 11.3 & 6.0e-17 & 9.9e-18 & \nodata & 0 & \nodata & \nodata & 0 \\
R20-106 & 13:36:56.385 & -29:48:17.78 & \nodata & \phn\phn289.9 $\pm$ \phn12.9 & \phn\phn130.8 $\pm$ \phn32.7 & \phn\phn284.2 $\pm$ \phn12.6 & 5.609 & 7 & 3.0 & 1.2 & 1.2 & 106 & \phn300.7 $\pm$ 11.7 & 4.0e-15 & 4.0e-16 & \nodata & 0 & \nodata & \nodata & 1 \\
R20-107 & 13:36:56.439 & -29:49:33.49 & \nodata & \phn\phn\phn50.1 $\pm$ \phn\phn8.2 & \phn\phn\phn58.4 $\pm$ \phn13.1 & \phn\phn\phn51.4 $\pm$ \phn\phn6.8 & 6.644 & 7 & 3.3 & 1.6 & 166.9 & 107 & \phn\phn43.5 $\pm$ \phn6.7 & 2.7e-14 & 3.9e-15 & M-25 & 11 & B12-108 & \nodata & 0 \\
R20-108 & 13:36:56.481 & -29:53:08.77 & \nodata & \phn\phn127.5 $\pm$ \phn\phn9.7 & \phn\phn\phn76.4 $\pm$ \phn11.7 & \phn\phn113.8 $\pm$ \phn\phn7.7 & 6.143 & 7 & 3.0 & 1.4 & 179.0 & 108 & \phn110.6 $\pm$ \phn7.1 & 1.5e-14 & 1.5e-15 & \nodata & 0 & \nodata & \nodata & 0 \\
R20-109 & 13:36:56.612 & -29:52:52.77 & \nodata & \phn\phn\phn65.3 $\pm$ \phn\phn8.5 & \phn\phn\phn43.5 $\pm$ \phn\phn9.8 & \phn\phn\phn56.6 $\pm$ \phn\phn6.3 & 6.434 & 7 & 2.8 & 1.7 & 172.7 & 109 & \phn\phn54.2 $\pm$ \phn6.2 & 6.7e-15 & 8.8e-16 & \nodata & 0 & \nodata & \nodata & 0 \\
R20-110 & 13:36:56.866 & -29:52:48.72 & \nodata & \phn\phn198.2 $\pm$ \phn12.1 & \phn\phn148.5 $\pm$ \phn13.1 & \phn\phn178.5 $\pm$ \phn\phn8.6 & 6.718 & 7 & 3.3 & 1.6 & 178.4 & 110 & \phn145.8 $\pm$ \phn5.6 & 5.4e-14 & 3.5e-15 & M-26 & 0 & \nodata & \nodata & 0 \\
R20-111 & 13:36:56.973 & -29:50:43.54 & Ea & \phn1673.9 $\pm$ \phn25.0 & \phn\phn822.4 $\pm$ \phn35.0 & \phn1561.5 $\pm$ \phn22.6 & 5.790 & 7 & 6.0 & 2.3 & 149.6 & 111 & \phn570.7 $\pm$ \phn6.4 & 3.5e-14 & 5.1e-16 & M-27 & 0 & \nodata & \nodata & 0 \\
R20-112 & 13:36:57.149 & -29:53:33.61 & \nodata & \phn\phn\phn59.1 $\pm$ \phn\phn8.4 & \phn\phn\phn12.6 $\pm$ \phn\phn8.8 & \phn\phn\phn59.0 $\pm$ \phn\phn8.4 & 5.503 & 7 & 2.7 & 1.4 & 178.5 & 112 & \phn\phn46.9 $\pm$ \phn7.0 & 2.1e-15 & 4.1e-16 & \nodata & 11 & B12-112 & \nodata & 0 \\
R20-113 & 13:36:57.369 & -29:50:35.25 & \nodata & \phn\phn\phn39.4 $\pm$ \phn\phn7.9 & \phn\phn\phn24.5 $\pm$ \phn10.9 & \phn\phn\phn36.6 $\pm$ \phn\phn6.6 & 6.131 & 7 & 3.0 & 1.2 & 178.2 & 113 & \phn\phn36.0 $\pm$ \phn4.7 & 6.1e-15 & 1.2e-15 & \nodata & 0 & \nodata & \nodata & 1 \\
R20-114 & 13:36:57.385 & -29:52:47.12 & \nodata & \phn\phn\phn92.0 $\pm$ \phn10.1 & \phn\phn\phn52.1 $\pm$ \phn11.2 & \phn\phn\phn80.6 $\pm$ \phn\phn7.8 & 6.273 & 7 & 3.1 & 1.4 & 8.7 & 114 & \phn\phn74.7 $\pm$ \phn5.5 & 2.0e-14 & 2.1e-15 & \nodata & 0 & \nodata & \nodata & 0 \\
R20-115 & 13:36:57.387 & -29:50:50.26 & Ea & \phn\phn253.1 $\pm$ \phn14.3 & \phn\phn\phn95.7 $\pm$ \phn19.1 & \phn\phn248.8 $\pm$ \phn14.1 & 5.549 & 7 & 5.3 & 2.1 & 139.8 & 115 & \phn\phn95.5 $\pm$ \phn6.5 & 8.8e-15 & 4.0e-16 & \nodata & 0 & \nodata & \nodata & 0 \\
R20-116 & 13:36:57.947 & -29:47:59.13 & Ec & \phn\phn\phn59.6 $\pm$ \phn\phn9.8 & \phn\phn\phn49.9 $\pm$ \phn28.8 & \phn\phn\phn58.4 $\pm$ \phn\phn9.5 & 5.645 & 7 & 3.1 & 1.5 & 1.1 & 116 & \phn\phn58.9 $\pm$ 10.0 & 2.5e-15 & 4.4e-16 & \nodata & 0 & \nodata & \nodata & 0 \\
R20-117 & 13:36:57.952 & -29:52:42.43 & \nodata & \phn\phn\phn80.4 $\pm$ \phn10.6 & \phn\phn\phn53.9 $\pm$ \phn16.5 & \phn\phn\phn76.0 $\pm$ \phn\phn9.0 & 6.144 & 7 & 9.3 & 4.1 & 37.0 & 117 & \phn\phn20.0 $\pm$ \phn3.9 & 1.3e-14 & 4.6e-16 & \nodata & 0 & \nodata & \nodata & 0 \\
R20-118 & 13:36:58.302 & -29:52:00.78 & \nodata & \phn\phn\phn51.0 $\pm$ \phn10.3 & \phn\phn\phn38.4 $\pm$ \phn11.0 & \phn\phn\phn46.0 $\pm$ \phn\phn8.0 & 6.451 & 7 & 3.2 & 1.7 & 6.9 & 118 & \phn\phn40.3 $\pm$ \phn6.2 & 9.8e-15 & 1.3e-15 & \nodata & 0 & \nodata & \nodata & 0 \\
R20-119 & 13:36:58.377 & -29:51:04.65 & Ea & \phn1133.8 $\pm$ \phn20.2 & \phn1141.1 $\pm$ \phn26.4 & \phn1137.2 $\pm$ \phn16.2 & 6.744 & 7 & 3.3 & 1.6 & 171.9 & 119 & 1152.4 $\pm$ 10.5 & 6.8e-15 & 2.9e-16 & M-28 & 0 & \nodata & \nodata & 1 \\
R20-120 & 13:36:58.639 & -29:52:36.94 & \nodata & \phn\phn\phn86.0 $\pm$ \phn12.0 & \phn\phn\phn27.7 $\pm$ \phn10.3 & \phn\phn\phn78.1 $\pm$ \phn10.6 & 5.749 & 7 & 2.7 & 1.2 & 0.4 & 120 & \phn\phn74.0 $\pm$ \phn8.5 & 2.8e-15 & 5.2e-16 & \nodata & 0 & \nodata & \nodata & 1 \\
R20-121 & 13:36:58.748 & -29:48:05.80 & \nodata & \phn\phn115.1 $\pm$ \phn12.9 & \phn\phn\phn54.7 $\pm$ \phn36.4 & \phn\phn113.6 $\pm$ \phn12.7 & 5.581 & 7 & 2.9 & 2.1 & 3.7 & 121 & \phn\phn75.8 $\pm$ 10.3 & 9.3e-14 & 5.7e-15 & \nodata & 0 & \nodata & \nodata & 0 \\
R20-122 & 13:36:58.967 & -29:51:26.59 & \nodata & \phn\phn\phn60.2 $\pm$ \phn13.2 & \phn\phn\phn56.0 $\pm$ \phn12.7 & \phn\phn\phn57.3 $\pm$ \phn\phn9.3 & 7.033 & 7 & 3.0 & 1.4 & 167.0 & 122 & \phn\phn57.3 $\pm$ \phn6.7 & 4.1e-14 & 6.0e-15 & \nodata & 0 & \nodata & \nodata & 0 \\
R20-123 & 13:36:58.985 & -29:50:25.78 & \nodata & \phn\phn\phn52.6 $\pm$ \phn\phn9.7 & \phn\phn\phn43.7 $\pm$ \phn11.4 & \phn\phn\phn48.0 $\pm$ \phn\phn7.5 & 6.386 & 7 & 3.5 & 1.2 & 169.5 & 123 & \phn\phn49.2 $\pm$ \phn5.0 & 3.2e-15 & 4.0e-16 & \nodata & 0 & \nodata & \nodata & 1 \\
R20-124 & 13:36:59.063 & -29:51:16.12 & Ea & \phn\phn623.7 $\pm$ \phn25.0 & \phn\phn245.8 $\pm$ \phn31.6 & \phn\phn602.2 $\pm$ \phn24.0 & 5.608 & 7 & 6.6 & 1.9 & 144.2 & 124 & \phn222.6 $\pm$ 11.3 & 2.3e-14 & 6.0e-16 & M-29 & 0 & \nodata & \nodata & 0 \\
R20-125 & 13:36:59.148 & -29:54:28.17 & \nodata & \phn\phn\phn62.2 $\pm$ \phn\phn9.9 & \phn\phn\phn50.8 $\pm$ \phn11.6 & \phn\phn\phn59.4 $\pm$ \phn\phn7.3 & 6.796 & 7 & 3.1 & 1.4 & 3.0 & 125 & \phn\phn59.0 $\pm$ \phn7.2 & 4.5e-14 & 6.3e-15 & \nodata & 0 & \nodata & \nodata & 1 \\
R20-126 & 13:36:59.174 & -29:51:48.23 & W & \phn\phn\phn34.1 $\pm$ \phn10.0 & \phn\phn\phn18.8 $\pm$ \phn10.0 & \phn\phn\phn29.8 $\pm$ \phn\phn7.5 & 6.394 & 7 & 2.6 & 1.4 & 174.1 & 126 & \phn\phn31.6 $\pm$ \phn5.6 & 3.2e-15 & 8.7e-16 & \nodata & 11 & B14-22 & \nodata & 1 \\
R20-127 & 13:36:59.189 & -29:52:31.02 & \nodata & \phn\phn\phn93.2 $\pm$ \phn14.0 & \phn\phn\phn65.7 $\pm$ \phn16.2 & \phn\phn\phn85.6 $\pm$ \phn11.1 & 6.437 & 7 & 7.3 & 5.4 & 32.2 & 127 & \phn\phn21.7 $\pm$ \phn4.2 & 1.3e-14 & 4.2e-16 & \nodata & 0 & \nodata & \nodata & 0 \\
R20-128 & 13:36:59.333 & -29:55:08.94 & \nodata & \phn\phn\phn37.2 $\pm$ \phn\phn8.2 & \phn\phn\phn24.5 $\pm$ \phn14.9 & \phn\phn\phn34.7 $\pm$ \phn\phn7.2 & 6.001 & 7 & 2.7 & 1.4 & 9.7 & 128 & \phn\phn40.2 $\pm$ \phn7.0 & 2.0e-15 & 5.4e-16 & \nodata & 11 & B12-122 & \nodata & 1 \\
R20-129 & 13:36:59.385 & -29:48:36.78 & \nodata & \phn\phn132.3 $\pm$ \phn12.4 & \phn\phn\phn91.7 $\pm$ \phn22.6 & \phn\phn125.7 $\pm$ \phn11.3 & 5.853 & 7 & 3.2 & 2.0 & 168.0 & 129 & \phn102.6 $\pm$ \phn8.2 & 3.3e-14 & 2.7e-15 & \nodata & 9 & B14-23 & B14-24 & 1 \\
R20-130 & 13:36:59.555 & -29:52:03.43 & \nodata & \phn\phn188.5 $\pm$ \phn18.8 & \phn\phn\phn71.6 $\pm$ \phn18.3 & \phn\phn170.9 $\pm$ \phn16.6 & 5.812 & 7 & 3.9 & 3.2 & 45.1 & 130 & \phn\phn91.5 $\pm$ 18.5 & 1.1e-14 & 1.4e-15 & \nodata & 11 & B12-124 & \nodata & 1 \\
R20-131 & 13:36:59.595 & -29:54:13.56 & W & \phn\phn\phn27.5 $\pm$ \phn\phn7.2 & \phn\phn\phn-1.2 $\pm$ \phn\phn9.6 & \phn\phn\phn43.1 $\pm$ \phn11.2 & 4.732 & 7 & 3.1 & 1.2 & 0.8 & 131 & \phn\phn21.5 $\pm$ \phn4.3 & 6.6e-16 & 1.9e-16 & \nodata & 0 & \nodata & \nodata & 1 \\
R20-132 & 13:36:59.657 & -29:48:37.18 & \nodata & \phn\phn\phn56.7 $\pm$ \phn10.3 & \phn\phn\phn21.5 $\pm$ \phn17.9 & \phn\phn\phn54.5 $\pm$ \phn\phn9.9 & 5.628 & 7 & 4.3 & 1.2 & 3.0 & 132 & \phn\phn52.0 $\pm$ \phn9.5 & 3.6e-14 & 6.1e-15 & \nodata & 0 & \nodata & \nodata & 0 \\
R20-133 & 13:36:59.823 & -29:52:25.63 & n & \phn\phn302.8 $\pm$ \phn22.7 & \phn\phn264.4 $\pm$ \phn20.4 & \phn\phn280.9 $\pm$ \phn15.2 & 7.114 & 7 & 4.6 & 1.5 & 2.9 & 133 & \phn215.2 $\pm$ 11.6 & 6.2e-14 & 3.5e-15 & \nodata & 0 & \nodata & \nodata & 2 \\
R20-134 & 13:36:59.825 & -29:48:43.24 & \nodata & \phn\phn\phn36.6 $\pm$ \phn\phn8.5 & \phn\phn\phn-0.9 $\pm$ \phn18.2 & \phn\phn\phn39.8 $\pm$ \phn\phn9.0 & 5.328 & 7 & 2.1 & 1.5 & 3.1 & 134 & \phn\phn38.7 $\pm$ \phn6.4 & 2.6e-14 & 6.6e-15 & \nodata & 0 & \nodata & \nodata & 0 \\
R20-135 & 13:36:59.956 & -29:51:04.37 & \nodata & \phn\phn\phn26.6 $\pm$ \phn\phn8.4 & \phn\phn\phn58.0 $\pm$ \phn15.4 & \phn\phn\phn42.5 $\pm$ \phn\phn8.3 & 7.500 & 7 & 4.7 & 4.1 & 38.4 & 135 & \phn\phn20.6 $\pm$ \phn4.0 & 6.1e-15 & 4.0e-16 & \nodata & 0 & \nodata & \nodata & 0 \\
R20-136 & 13:37:00.050 & -29:52:16.44 & n & \phn3251.0 $\pm$ \phn56.5 & \phn2196.3 $\pm$ \phn61.3 & \phn2908.8 $\pm$ \phn44.3 & 6.328 & 7 & 6.5 & 5.0 & 9.5 & 136 & \phn818.6 $\pm$ 18.8 & 2.6e-13 & 2.3e-15 & M-31 & 3 & B14-28 & \nodata & 7 \\
R20-137 & 13:37:00.089 & -29:48:40.86 & W & \phn\phn\phn25.3 $\pm$ \phn\phn9.4 & \phn\phn\phn31.1 $\pm$ \phn16.6 & \phn\phn\phn25.3 $\pm$ \phn\phn7.6 & 6.606 & 7 & 2.5 & 1.6 & 8.5 & 137 & \phn\phn27.2 $\pm$ \phn5.8 & 6.8e-15 & 1.9e-15 & \nodata & 11 & B12-130 & \nodata & 0 \\
R20-138 & 13:37:00.120 & -29:52:02.06 & n & \phn4812.8 $\pm$ \phn57.0 & \phn2748.9 $\pm$ \phn49.0 & \phn4049.9 $\pm$ \phn40.9 & 6.452 & 7 & 7.8 & 4.3 & 170.6 & 138 & 1215.8 $\pm$ 34.8 & 2.1e-13 & 3.5e-15 & \nodata & 9 & D10-N-03 & B14-27 & 6 \\
R20-139 & 13:37:00.200 & -29:51:37.18 & n & \phn1653.6 $\pm$ \phn44.4 & \phn1217.1 $\pm$ \phn40.9 & \phn1444.4 $\pm$ \phn31.2 & 6.678 & 7 & 7.6 & 6.0 & 173.9 & 139 & \phn310.9 $\pm$ 21.7 & 1.0e-13 & 9.8e-16 & \nodata & 0 & \nodata & \nodata & 1 \\
R20-140 & 13:37:00.320 & -29:51:51.13 & n & 49874.2 $\pm$ 301.0 & 36160.9 $\pm$ 263.6 & 43626.4 $\pm$ 206.3 & 6.840 & 7 & 8.6 & 5.5 & 11.1 & 140 & 8133.2 $\pm$ 72.5 & 3.0e-12 & 1.1e-14 & \nodata & 11 & D10-N-02 & \nodata & 13 \\
R20-141 & 13:37:00.414 & -29:52:00.57 & n & \phn9981.5 $\pm$ 115.7 & \phn7491.5 $\pm$ \phn99.4 & \phn8800.3 $\pm$ \phn77.4 & 6.921 & 7 & 6.7 & 3.6 & 166.2 & 141 & 3508.6 $\pm$ 94.5 & 1.4e-12 & 4.8e-14 & \nodata & 9 & D10-N-05 & D10-N-06 & 5 \\
R20-142 & 13:37:00.424 & -29:52:22.63 & \nodata & \phn\phn182.5 $\pm$ \phn17.1 & \phn\phn\phn94.6 $\pm$ \phn13.3 & \phn\phn148.4 $\pm$ \phn11.9 & 6.509 & 7 & 3.1 & 1.2 & 0.6 & 142 & \phn153.7 $\pm$ 12.9 & 1.0e-14 & 1.4e-15 & \nodata & 11 & B14-31 & \nodata & 1 \\
R20-143 & 13:37:00.586 & -29:54:18.95 & \nodata & \phn\phn147.0 $\pm$ \phn12.0 & \phn\phn119.3 $\pm$ \phn14.4 & \phn\phn140.0 $\pm$ \phn\phn9.2 & 6.736 & 7 & 3.2 & 1.5 & 5.4 & 143 & \phn126.6 $\pm$ \phn7.2 & 8.1e-14 & 6.7e-15 & M-33 & 0 & \nodata & \nodata & 0 \\
\hline
\multicolumn{21}{l}{(This table is available in its entirety in machine-readable form.)}
\end{tabular}
\end{center}
\end{sidewaystable*}

In our deep ATCA radio observations of M83, we detect 266 radio sources at a significance of $4\sigma$ or above and an additional 4 sources below $4\sigma$ but with associations with SNRs or X-ray sources (Figure~\ref{fig_overview} right). A sample of the radio catalogue is presented in Table~\ref{tab:radio_catalogue}, where the complete table is supplied online. While this is not the first radio study of M83 \citep[e.g.,][]{maddox06}, our ATCA observations provide the most sensitive catalogue of M83 to date, with a typical rms noise at 7.25~GHz of $\sim6$\,\uJybeam. At the distance of M83, this corresponds to a 1-$\sigma$ 7.25~GHz radio luminosity sensitivity limit of $1.1 \times 10^{33}\,(D/4.61\, \mathrm{Mpc})^2$ \,erg\,s$^{-1}$. These radio observations provide high spatial resolution, with angular resolution of a few arcsec (see Section~\ref{sec:ATCA} for full details).

Table~\ref{tab:radio_cat_columns} describes the columns in the radio catalogue.  There are several columns of particular interest: the \textit{Wrn} flag both indicates sources with possible issues and also identifies likely extragalactic sources (behind the galaxy); there are two columns with information on \ha\ emission associated with the object, computed by integrating the continuum-subtracted Magellan \ha\ image from \citet{blair12} over the island associated with the object.\footnote{Most \ha\ imaging filters also pass some fraction of \NiiL, depending on exact filter widths and velocity shifts that may be present. The ratio of [N~II]:\ha\ also changes between photoionized and shocked regions, and as a function of galactic radius.  For the Magellan and {\it HST} observations, this contamination is generally small, and here and elsewhere we simply refer to ``\ha'' fluxes.}
The  \textit{AltName}, \textit{Sflag}, \textit{SNRname} and \textit{nXray} columns indicate associations with various other catalogs.

Table~\ref{tab:flags} provides additional information on the \textit{Sflag} column, which provides information on the reliability of associations
between the radio sources and SNRs.

\subsection{Radio Spectral Indices}
\label{sec:spectralindices}

Radio emission from spiral galaxies typically arises from thermal (free-free) and non-thermal (synchrotron) emission, where the thermal emission arises mostly from \hii\ regions, while the non-thermal comes mostly from SNRs and, in a few cases, from hot spots and lobes powered by jets from accreting black holes (BHs) and neutron stars. The thermal or non-thermal nature of a source is characterized by its spectral index, $\alpha$, defined as $F_\nu(\alpha) \propto \nu^\alpha$.
Therefore, a population of radio sources in a galaxy typically exhibits both steep ($\alpha<0$) and flat or inverted ($\alpha \gtrsim 0$) spectral indices: 
flatter spectral indices are expected for free-free emission, while steeper spectra are expected for synchrotron emission. For a typical radio source population, the expected distribution will be bi-modal with peaks at around $\alpha \approx -0.5$ and $\alpha \approx 0$, with some amount of overlap  \citep[e.g., for M33 see Figure~13 of][]{white19}.

\begin{figure}
\centering
\includegraphics[width=0.95\columnwidth]{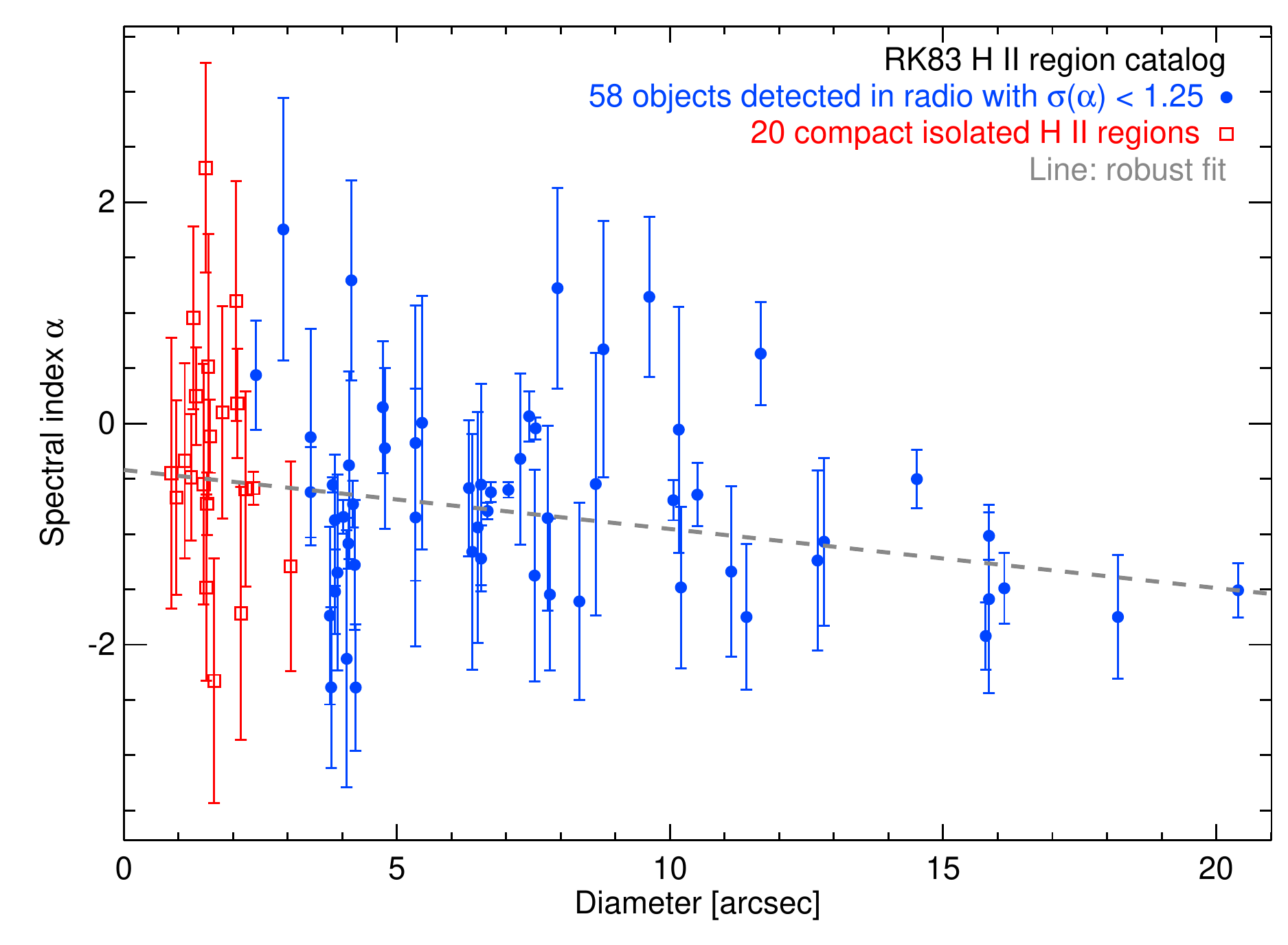}
\caption{
	Variation of fitted spectral index with size for a sample of \hii\ regions.  The larger objects come from \citet{rumstay83}, while the smaller objects are a sample of compact, isolated regions selected visually from our \ha\ image.  All of these should be sources of optically thin, free-free radio emission with a spectral index $\alpha \approx -0.1$.  There is a clear shift toward steeper spectral indices as the sizes increase. This is probably attributable to missing flux in the higher resolution 9 GHz due to limitations in the ATCA visibility coverage.  Based on this and other tests of the data, we have concluded that spectral indices derived from these data are not reliable, and we do not use them for analysis in this paper.
\label{fig:si_vs_size}}
\end{figure}

While we present both 5.5 and 9~GHz flux densities, we do not believe the spectral indices from our data are reliable and caution against their use.
Analysis of the population  of detected radio sources in M83 shows steeper than expected radio spectral indices, and there is not a large population of flat-spectrum thermal sources (even though many sources align with optical \hii\ regions).
The measured spectral index distribution is very different from that observed in M33 \citep{white19}. In M83, our results unexpectedly show steep radio spectra for sources identified from optical and X-ray studies as \hii\ regions, and the spectral indices are size-dependent (albeit with significant scatter; Figure~\ref{fig:si_vs_size}). By contrast in M33, the spectral indices of well-detected sources showed a bimodal distribution with the flat-spectrum sources associated preferentially with \hii\ regions and with the steep-spectrum sources associated with SNRs or X-ray sources, most of which are likely background AGN.

An extensive exploration of our data revealed no signs of observational issues such as phase decorrelation (Section~\ref{sec:ATCA}). While we do not have a definitive explanation for the missing flat-spectrum population, we believe this is likely due to the systematic loss of extended radio emission in the higher frequency, hence higher resolution, observations.
The point-spread function area varies by a factor of 5 from the low-frequency edge of the 5.5~GHz bandpass to the high-frequency edge of the 9~GHz bandpass. Consequently, sources that are only slightly resolved at low frequencies are quite over-resolved at high frequencies, with the attendant suppression of the high-frequency flux measurements. That, in turn, naturally leads to steepening of the fitted spectral indices.\footnote{Note that the M33 VLA observations analyzed by \citet{white19} used 1.45~GHz and 5~GHz scaled-array observations, with matched PSFs in the two bandpasses, reducing resolution biases.} This effect is shown in Figure~\ref{fig:si_vs_size} for the \hii\ region sample, where the objects are expected to have spectral indices near the free-free value $\alpha = -0.1$.
Therefore, we conclude that the spectral indices derived from fitting the 5.5 and 9~GHz flux densities are not useful to discriminate source populations, and we were unable to correct the source-dependent biases. We have chosen not to include the $\alpha$ values in the catalogue, and we do not rely on them for any analysis in this paper.

Before leaving this topic, we note that there have been a number of attempts to identify SNRs based on their radio properties.  Most of these have achieved only a limited degree of success because of limitations associated with the sensitivity and spatial resolution of the instrumentation, but that is changing as instrumentation has improved.   At radio wavelengths, SNRs are typically identified as  non-thermal radio sources that, to reduce the possibility of contamination by background sources, are  associated with \ha\ emission.  We had hoped to use the ATCA data for this purpose.  As it is, however, we failed in this objective, because of our inability to derive accurate spectral indices.  Even had we succeeded, in the sense that the spectral indices had been less biased toward steeper values, we would still have had to deal with the fact that at the brightnesses of the radio SNRs in M83 and the sensitivity of our survey, the statistical errors on the spectral indices would have been substantial for most of the sources.  Given the fact that one of the criteria for declaring a source a SNR candidate is that it be associated with \ha\ emission and that most \ha\ emission in galaxies arises not from SNRs but from \hii\ regions, which are expected to produce radio bremsstrahlung emission, very high quality data is required to confidently establish a radio source as a candidate for a radio SNR.  One possibility is to do a good job in subtracting off the thermal emission, but this requires accurately fluxed \ha\ images.  We feel this has not been sufficiently appreciated in establishing a number of the existing samples of radio SNRs.  That said, we believe it is important to obtain the data necessary to really begin to identify SNRs reliably at radio wavelengths, since unless this can be done, the sample of SNRs in external galaxies will continue to be biased by the optical properties of SNRs.

\begin{table*}
\caption{Radio Source Catalog Column Descriptions\label{tab:radio_cat_columns}}
\begin{center}
\begin{tabular}{lcl}
\hline
\hline
Column Name & Units & Description \\
\hline
Name         &        & Name of source (= R20-island number) \\
RA           & hh:mm:ss.sss & J2000 RA (flux-weighted centroid) \\
Dec          & ${+}$dd:mm:ss.ss & J2000 Declination (flux-weighted centroid \\
RAPeak       & deg    & J2000 RA (peak) \\
DecPeak      & deg    & J2000 Declination (peak) \\
Wrn          &        & n = in confused nuclear region, W = below detection threshold, E* = extragalactic candidate$^a$ \\
F5.5GHz      & \uJy   & Integrated flux density at frequency 5.5 GHz \\
F5.5Err      & \uJy   & Error on F5.5GHz \\
F9GHz        & \uJy   & Integrated flux density at frequency 9.2 GHz \\
F9Err        & \uJy   & Error on F9GHz \\
Fint         & \uJy   & Flux density at pivot frequency pFreq, $\Fint$ \\
FiErr        & \uJy   & Error on Fint \\
pFreq        & GHz    & Pivot frequency $\nu_0$ where signal-to-noise is maximized \\
nBands       &        & Number of frequency bands with data (1 to 7 of $4\times5.5$, $3\times9$ GHz subbands) \\
Major        & arcsec & Major axis FWHM (includes Gaussian beam with FWHM = $3.2\times1.4$ arcsec) \\
Minor        & arcsec & Minor axis FWHM (includes Gaussian beam with FWHM = $3.2\times1.4$ arcsec) \\
PA           & deg    & Position angle of major axis \\
Island       &        & Island number \\
Fpeak        & \uJy\,beam$^{-1}$   & Flux density in peak pixel in island from detection image, $F_p$ \\
FpErr        & \uJy\,beam$^{-1}$   & Noise in Fpeak \\
$H\alpha(\mathrm{tot})$ & $\mathrm{erg\, cm^{-2} s^{-1}}$ & Integrated $H\alpha$ flux$^b$ \\
$H\alpha(\mathrm{ave})$ & $\mathrm{erg\, cm^{-2} s^{-1} {\rm arcsec}^{-2}}$ & $H\alpha$ surface brightness \\
AltName      &        & Matching source from \citet{maddox06} and historical SNe \\
Sflag        &        & SNR detection flag$^c$ \\
SNRname[1-3] &        & Names of associated SNR(s)$^d$ \\
nXray        &        & Number of associated X-ray sources in \cite{long14} \\
\hline
\multicolumn{3}{l}{
\begin{minipage}{0.9\textwidth}
Column descriptions for Table~\ref{tab:radio_catalogue}. \\
$^a$ 266 sources are detected at $4\sigma$ or greater.
4 sources marked as 'W' are below the detection threshold but have SNR or X-ray counterparts.
14 sources are in the confused nuclear region where identification of counterparts
is very difficult. The extragalactic flag has values Ea (clear optical galaxy or radio double),
Eb (stellar counterpart without H$\alpha$ emission), Ec (no counterpart in optical, but also no
association with either structures or dust in M83). A `?' indicates greater uncertainty in the E classification.\\
$^b$ The $H\alpha$ fluxes were obtained from images described by \cite{blair12} and include some contribution
from [N~II]$\lambda\lambda$6548, 6583.\\
$^c$ The detection flag (Sflag) is a bit flag where the bits indicate
whether the association between the radio source and the SNR is unambiguous:
1 = this source may have a match in the SNR catalog;
2 = unambiguous match: this source matches only one object from the SNR catalog;
8 = mutually good match: this source is best for the other object, and the other object is best for this source.
The most reliable matches will have flag bit 8 set.  All of those will have bit 1 set as well, and most of them
will also have bit 2 set.
See Table~\ref{tab:flags} for more information.\\
$^d$ SNRs are taken from \cite{dopita10}, \cite{blair12}, or \cite{blair14}.
	Multiple SNR matches are listed in order of decreasing island overlap.\\
\end{minipage}}
\end{tabular}
\end{center}
\end{table*}

\begin{table}
\caption{Flag Value Counts in the Radio and Forced Photometry Catalogs \label{tab:flags}}
\begin{tabular}{crrrl}
\hline
	Value &
	\pb{6ex}{Sflag\\Radio$^a$} &
	\pb{6ex}{Rflag\\SNR$^b$} &
	\pb{7ex}{Rflag\\X-ray$^c$} &
	Meaning
\\
\hline
\hline
0  & 187 & 200 & 292 & \pb{10em}{No match} \\
1  &   1 &   7 &  19 & \pb{10em}{Ambiguous matches} \\
3  &   3 &  18 &  20 & \pb{10em}{Single match but not mutually good} \\
9  &  14 &   3 &   6 & \pb{10em}{Mutually good match} \\
11 &  65 &  76 &  87 & \pb{10em}{Single, mutually good match} \\
\hline
\multicolumn{5}{l}{
\begin{minipage}{0.9\columnwidth}
$^a$ The Sflag column in the radio catalog indicates the quality of SNR associations.\\
$^b$ The Rflag column in the SNR forced photometry catalog indicates the quality of radio source associations.\\
$^c$ The Rflag column in the X-ray forced photometry catalog indicates the quality of radio source associations.\\
\end{minipage}}
\end{tabular}
\end{table}

\subsection{Forced Photometry Catalogues}
\label{sec:forced_photometry}

The radio catalogue described above was constructed without any reference to the positions of known
optical SNRs or other sources in M83 (although many of the remnants were detected).  We have also
constructed separate catalogues of radio properties at the positions of known objects by
integrating the radio images over the elliptical region determined for each optical SNR \citep{williams19} and
X-ray source \citep{long14}.
This method allowed us to establish either measured flux densities
or appropriate upper limits for the radio emission from these sources, whether or not a
radio source was independently detected at that position.  We will refer to these versions of the radio
data as the ``forced photometry'' SNR or X-ray extraction catalogues.

The calculations of flux densities and flux-weighted positions proceeds much as
described above for the radio catalogue \citep[for more details see][]{white19}.  The island map for the forced photometry catalogue was
determined from the elliptical regions in the external catalogue rather than from the radio map itself.
Since the island map makes no reference to the radio morphology, we did not have multi-resolution
islands but instead summed the radio flux density over the entire region using the single background-subtracted
image in each radio band.
The region sizes were expanded by convolving them with the elliptical radio beam shape; this led to a more
representative sampling of the lower-resolution radio image, particularly for {\it HST}-discovered SNRs that
can be very small in angular size.

Associations between SNRs, X-ray sources, and objects in our master radio catalogue were determined by the overlap between
the radio island map and the forced-photometry island maps.  Many objects have unambiguous matches with catalogued
radio sources, but there are also ambiguous cases where, for example, a single radio island overlaps several SNR
islands or vice versa.  The information on the overlapping sources and a flag that captures information
on the ambiguity of the association are also included in the tables.

In the discussions that follow, we use 3$\sigma$ as the detection limit in the forced photometry catalogues, as we did in \citet{white19}.  We use this lower limit than is used in the main catalogue, because in the forced photometry catalogue the position of the source is fixed.
Samples of the SNR forced photometry catalogue is found in Table~\ref{tab:forced_snr} and the X-ray catalogue in Table~\ref{tab:forced_xray}, where the full catalogues are provided online.
Table~\ref{tab:forced_columns} describes the columns in the SNR and X-ray forced photometry catalogues.  The columns are similar to those in the radio catalogue (see Table~\ref{tab:radio_cat_columns}) but there are additional columns giving the galactocentric distance $\rho$ (in kpc), the SNR diameter $D$ (pc), and information on associations with objects in the main radio catalogue.

\begin{table*}
\setlength{\tabcolsep}{1pt}
\scriptsize
\begin{center}
\caption{Forced Photometry of M83 SNR Candidates \label{tab:forced_snr}}
\begin{tabular}{cccrrcccccccccccccc}
\hline
\hline
Name & RA & Dec & $\rho$ & D & Cl & Wrn & F5.5GHz & F9GHz & $\Fint$ & $\nu_p$ & nBands & Island & H$\alpha(\mathrm{tot})$ & H$\alpha(\mathrm{ave})$ & Rflg & RadName1 & RadName2 & nXray \\
~ & (J2000) & (J2000) & (kpc) & (pc) & ~ & ~ & ($\mu$Jy) & ($\mu$Jy) & ($\mu$Jy) & (GHz) & ~ & ~ & \pbox{5em}{\centering $(\mathrm{erg}\,\mathrm{cm}^{-2}$\break$\mathrm{s}^{-1})$} & \pbox{5em}{\centering $(\mathrm{erg}\,\mathrm{cm}^{-2}$\break$\mathrm{s}^{-1}\mathrm{arcsec}^{-2})$} & ~ & ~ & ~ & ~ \\
\hline
B14-20 & 13:36:58.899 & -29:52:26.26 & 0.91 & 13.0 & C & \nodata & \phn28.3 $\pm$ 13.5 & -17.1 $\pm$ 13.1 & \phn44.4 $\pm$ 21.2 & 4.732 & 7 & 1 & 1.3e-15 & 4.8e-16 & 0 & \nodata & \nodata & 1 \\
B14-22 & 13:36:59.169 & -29:51:47.90 & 0.59 & 5.8 & A & \nodata & \phn12.2 $\pm$ 13.1 & \phn23.7 $\pm$ 13.1 & \phn24.5 $\pm$ 10.4 & 8.623 & 7 & 2 & 2.1e-15 & 8.5e-16 & 11 & R20-126 & \nodata & 1 \\
B14-28 & 13:37:00.065 & -29:52:08.75 & 0.39 & 14.8 & C & n & 174.0 $\pm$ 23.3 & \phn29.2 $\pm$ 25.2 & 180.2 $\pm$ 24.0 & 5.427 & 7 & 3 & 4.9e-15 & 1.6e-15 & 1 & R20-146 & R20-138 & 0 \\
D10-17 & 13:37:04.877 & -29:52:18.58 & 1.36 & 20.6 & C & \nodata & \phn\phn9.2 $\pm$ \phn7.1 & \phn-1.8 $\pm$ \phn9.8 & \phn\phn9.3 $\pm$ \phn7.1 & 5.485 & 7 & 5 & 1.2e-15 & 3.8e-16 & 0 & \nodata & \nodata & 0 \\
B14-51 & 13:37:06.989 & -29:51:09.59 & 2.05 & 10.7 & D & \nodata & \phn-3.0 $\pm$ \phn7.5 & \phn\phn5.7 $\pm$ 10.9 & \phn-4.8 $\pm$ 11.8 & 4.732 & 7 & 6 & 2.6e-15 & 9.3e-16 & 0 & \nodata & \nodata & 0 \\
B14-54 & 13:37:08.331 & -29:50:56.33 & 2.53 & 25.5 & D & \nodata & \phn10.1 $\pm$ \phn7.0 & \phn-0.8 $\pm$ \phn9.2 & \phn15.8 $\pm$ 11.0 & 4.732 & 7 & 7 & 1.1e-14 & 2.9e-15 & 0 & \nodata & \nodata & 1 \\
B14-58 & 13:37:09.315 & -29:50:58.53 & 2.77 & 20.1 & C & \nodata & \phn37.0 $\pm$ \phn8.4 & \phn46.0 $\pm$ 11.2 & \phn38.1 $\pm$ \phn6.4 & 6.951 & 7 & 8 & 2.6e-14 & 7.5e-15 & 11 & R20-228 & \nodata & 0 \\
D10-38 & 13:37:10.364 & -29:51:33.93 & 2.87 & 25.0 & Af & \nodata & \phn17.0 $\pm$ \phn8.1 & -14.7 $\pm$ 12.5 & \phn26.7 $\pm$ 12.7 & 4.732 & 7 & 9 & 5.1e-15 & 1.3e-15 & 0 & \nodata & \nodata & 0 \\
D10-40 & 13:37:10.839 & -29:52:44.47 & 3.32 & 8.7 & Af & \nodata & \phn15.6 $\pm$ \phn6.5 & \phn14.4 $\pm$ 10.9 & \phn24.5 $\pm$ 10.3 & 4.732 & 7 & 10 & 1.0e-15 & 3.7e-16 & 0 & \nodata & \nodata & 0 \\
B14-03 & 13:36:50.116 & -29:52:43.67 & 3.39 & 14.3 & C & \nodata & 211.9 $\pm$ \phn7.7 & 124.8 $\pm$ 12.4 & 198.3 $\pm$ \phn6.8 & 5.940 & 7 & 30 & 7.7e-15 & 2.7e-15 & 11 & R20-034 & \nodata & 0 \\
B14-07 & 13:36:51.192 & -29:50:42.32 & 3.56 & 5.4 & B & \nodata & 237.5 $\pm$ \phn8.0 & 130.6 $\pm$ 11.5 & 221.6 $\pm$ \phn7.1 & 5.927 & 7 & 31 & 9.9e-15 & 3.9e-15 & 11 & R20-043 & \nodata & 1 \\
B14-08 & 13:36:51.483 & -29:52:33.24 & 2.94 & 25.0 & A & \nodata & \phn21.8 $\pm$ \phn8.0 & \phn\phn9.7 $\pm$ 13.1 & \phn34.3 $\pm$ 12.6 & 4.732 & 7 & 32 & 3.9e-15 & 1.0e-15 & 0 & \nodata & \nodata & 0 \\
B14-09 & 13:36:51.526 & -29:53:00.92 & 3.13 & 16.1 & C & \nodata & 235.3 $\pm$ \phn8.2 & 164.5 $\pm$ 13.4 & 223.3 $\pm$ \phn7.2 & 5.994 & 7 & 33 & 1.5e-14 & 5.0e-15 & 11 & R20-048 & \nodata & 0 \\
B14-10 & 13:36:51.806 & -29:52:01.90 & 2.78 & 10.3 & D & \nodata & -19.7 $\pm$ \phn8.6 & -27.0 $\pm$ 12.6 & -24.3 $\pm$ \phn6.8 & 7.351 & 7 & 34 & 8.4e-16 & 3.0e-16 & 0 & \nodata & \nodata & 0 \\
B14-13 & 13:36:53.729 & -29:48:51.26 & 5.02 & 34.0 & C & \nodata & \phn67.6 $\pm$ \phn9.5 & \phn49.9 $\pm$ 21.4 & \phn66.3 $\pm$ \phn8.3 & 6.074 & 7 & 35 & 3.7e-14 & 7.8e-15 & 11 & R20-079 & \nodata & 0 \\
B14-19 & 13:36:58.643 & -29:51:06.49 & 1.40 & 4.0 & C & \nodata & \phn\phn3.3 $\pm$ 11.5 & \phn\phn2.2 $\pm$ 15.8 & \phn\phn5.2 $\pm$ 18.1 & 4.732 & 7 & 36 & 1.4e-15 & 5.4e-16 & 0 & \nodata & \nodata & 1 \\
B14-23 & 13:36:59.316 & -29:48:36.51 & 4.73 & 14.3 & C & \nodata & \phn83.1 $\pm$ \phn8.9 & \phn61.3 $\pm$ 18.2 & \phn80.1 $\pm$ \phn8.3 & 5.801 & 7 & 37 & 6.6e-15 & 2.4e-15 & 11 & R20-129 & \nodata & 1 \\
B14-24 & 13:36:59.442 & -29:48:36.99 & 4.72 & 12.5 & C & \nodata & \phn97.1 $\pm$ \phn9.3 & \phn37.3 $\pm$ 19.4 & \phn95.2 $\pm$ \phn9.1 & 5.577 & 7 & 38 & 8.2e-15 & 3.0e-15 & 3 & R20-129 & \nodata & 1 \\
B14-25 & 13:36:59.789 & -29:48:37.87 & 4.68 & 10.7 & C & \nodata & \phn52.4 $\pm$ 10.2 & \phn30.6 $\pm$ 20.7 & \phn57.6 $\pm$ 10.8 & 5.286 & 7 & 39 & 1.4e-14 & 5.0e-15 & 0 & \nodata & \nodata & 0 \\
B14-31 & 13:37:00.415 & -29:52:22.55 & 0.64 & 14.3 & B & \nodata & 195.5 $\pm$ 18.0 & 132.5 $\pm$ 15.3 & 166.8 $\pm$ 12.1 & 6.783 & 7 & 40 & 4.3e-15 & 1.4e-15 & 11 & R20-142 & \nodata & 1 \\
\hline
\multicolumn{19}{l}{(This table is available in its entirety in machine-readable form.)}
\end{tabular}
\end{center}
\end{table*}

\begin{table*}
\setlength{\tabcolsep}{1pt}
\scriptsize
\begin{center}
\caption{Forced Photometry of M83 X-ray Sources \label{tab:forced_xray}}
\begin{tabular}{cccrcccccccccccccc}
\hline
\hline
Name & RA & Dec & $\rho$ & Cl & Wrn & F5.5GHz & F9GHz & $\Fint$ & $\nu_p$ & nBands & Island & H$\alpha(\mathrm{tot})$ & H$\alpha(\mathrm{ave})$ & Rflg & RadName1 & RadName2 & nSNR \\
~ & (J2000) & (J2000) & (kpc) & ~ & ~ & ($\mu$Jy) & ($\mu$Jy) & ($\mu$Jy) & (GHz) & ~ & ~ & \pbox{5em}{\centering $(\mathrm{erg}\,\mathrm{cm}^{-2}$\break$\mathrm{s}^{-1})$} & \pbox{5em}{\centering $(\mathrm{erg}\,\mathrm{cm}^{-2}$\break$\mathrm{s}^{-1}\mathrm{arcsec}^{-2})$} & ~ & ~ & ~ & ~ \\
\hline
X169 & 13:36:58.629 & -29:52:36.76 & 1.15 & A & \nodata & \phn82.8 $\pm$ \phn15.2 & \phn43.1 $\pm$ 14.7 & \phn71.8 $\pm$ \phn12.0 & 6.158 & 7 & 169 & 3.2e-15 & 4.9e-16 & 11 & R20-120 & \nodata & 0 \\
X170 & 13:36:58.656 & -29:51:06.69 & 1.39 & D & \nodata & \phn\phn4.2 $\pm$ \phn13.4 & \phn\phn5.5 $\pm$ 18.7 & \phn\phn6.7 $\pm$ \phn21.0 & 4.732 & 7 & 170 & 3.3e-15 & 5.4e-16 & 0 & \nodata & \nodata & 1 \\
X171 & 13:36:58.668 & -29:43:36.40 & 11.78 & D & \nodata & \phn51.7 $\pm$ 113.1 & \nodata & \phn51.7 $\pm$ 113.1 & 4.732 & 1 & 171 & 0.0e+00 & 0.0e+00 & 0 & \nodata & \nodata & 0 \\
X172 & 13:36:58.713 & -29:51:00.60 & 1.51 & A & \nodata & \phn38.4 $\pm$ \phn13.6 & \phn37.2 $\pm$ 17.6 & \phn38.4 $\pm$ \phn11.1 & 6.707 & 7 & 172 & 4.5e-15 & 7.3e-16 & 0 & \nodata & \nodata & 1 \\
X173 & 13:36:58.800 & -29:48:31.74 & 4.88 & D & \nodata & \phn-4.6 $\pm$ \phn\phn4.2 & -29.5 $\pm$ 21.9 & -25.8 $\pm$ \phn23.3 & 9.768 & 7 & 173 & 2.1e-15 & 3.4e-16 & 0 & \nodata & \nodata & 0 \\
X174 & 13:36:58.812 & -29:44:33.28 & 10.44 & D & \nodata & -10.0 $\pm$ \phn83.3 & \nodata & -55.9 $\pm$ 466.8 & 9.768 & 2 & 174 & -4.0e-16 & -6.4e-17 & 0 & \nodata & \nodata & 0 \\
X175 & 13:36:58.908 & -29:50:38.72 & 1.96 & D & \nodata & -12.5 $\pm$ \phn10.6 & \phn\phn3.3 $\pm$ 13.9 & -19.7 $\pm$ \phn16.6 & 4.732 & 7 & 175 & 1.7e-15 & 2.7e-16 & 0 & \nodata & \nodata & 0 \\
X176 & 13:36:58.912 & -29:52:25.60 & 0.90 & B & \nodata & \phn29.1 $\pm$ \phn15.7 & -13.4 $\pm$ 15.7 & \phn45.6 $\pm$ \phn24.6 & 4.732 & 7 & 176 & 3.0e-15 & 4.9e-16 & 0 & \nodata & \nodata & 1 \\
X177 & 13:36:58.956 & -29:50:24.96 & 2.26 & A & \nodata & \phn35.5 $\pm$ \phn10.3 & \phn37.7 $\pm$ 13.9 & \phn35.0 $\pm$ \phn\phn8.2 & 6.621 & 7 & 177 & 2.6e-15 & 4.1e-16 & 11 & R20-123 & \nodata & 0 \\
X178 & 13:36:59.090 & -29:53:36.20 & 2.38 & B & \nodata & \phn14.5 $\pm$ \phn10.2 & \phn\phn8.5 $\pm$ 14.8 & \phn13.6 $\pm$ \phn\phn8.8 & 6.129 & 7 & 178 & 3.4e-15 & 5.2e-16 & 0 & \nodata & \nodata & 0 \\
X179 & 13:36:59.114 & -29:54:27.68 & 3.56 & A & \nodata & \phn71.1 $\pm$ \phn10.3 & \phn77.2 $\pm$ 14.3 & \phn73.9 $\pm$ \phn\phn8.3 & 6.812 & 7 & 179 & 4.0e-14 & 6.4e-15 & 11 & R20-125 & \nodata & 0 \\
X180 & 13:36:59.186 & -29:51:44.78 & 0.61 & D & \nodata & -30.2 $\pm$ \phn15.7 & -17.6 $\pm$ 15.4 & -26.3 $\pm$ \phn12.2 & 6.303 & 7 & 180 & 4.2e-15 & 6.9e-16 & 0 & \nodata & \nodata & 0 \\
X181 & 13:36:59.186 & -29:51:48.02 & 0.58 & A & \nodata & \phn\phn4.3 $\pm$ \phn\phn3.0 & \phn17.6 $\pm$ 15.3 & \phn24.2 $\pm$ \phn16.8 & 9.768 & 7 & 181 & 5.3e-15 & 8.6e-16 & 11 & R20-126 & \nodata & 1 \\
X182 & 13:36:59.318 & -29:53:17.55 & 1.94 & D & \nodata & \phn-8.5 $\pm$ \phn10.8 & -28.8 $\pm$ 17.3 & -21.4 $\pm$ \phn12.3 & 8.378 & 7 & 182 & 1.9e-15 & 3.0e-16 & 0 & \nodata & \nodata & 0 \\
X183 & 13:36:59.332 & -29:55:08.82 & 4.51 & A & \nodata & \phn43.2 $\pm$ \phn\phn9.7 & \phn23.7 $\pm$ 18.9 & \phn41.6 $\pm$ \phn\phn9.2 & 5.702 & 7 & 183 & 3.0e-15 & 5.0e-16 & 11 & R20-128 & \nodata & 1 \\
X184 & 13:36:59.361 & -29:48:37.43 & 4.71 & B & \nodata & 120.8 $\pm$ \phn10.3 & \phn68.5 $\pm$ 21.2 & 116.9 $\pm$ \phn\phn9.9 & 5.674 & 7 & 184 & 1.9e-14 & 3.1e-15 & 11 & R20-129 & \nodata & 3 \\
X185 & 13:36:59.455 & -29:49:59.04 & 2.81 & D & \nodata & -10.8 $\pm$ \phn\phn7.5 & -13.4 $\pm$ 13.3 & -16.9 $\pm$ \phn11.8 & 4.732 & 7 & 185 & 2.0e-15 & 3.2e-16 & 0 & \nodata & \nodata & 0 \\
X186 & 13:36:59.508 & -29:52:04.03 & 0.47 & B & \nodata & 197.5 $\pm$ \phn21.8 & \phn54.8 $\pm$ 23.0 & 212.2 $\pm$ \phn23.1 & 5.356 & 7 & 186 & 6.8e-15 & 1.1e-15 & 11 & R20-130 & \nodata & 1 \\
X187 & 13:36:59.584 & -29:54:13.74 & 3.23 & A & \nodata & \phn26.1 $\pm$ \phn\phn8.8 & \phn16.1 $\pm$ 14.1 & \phn26.0 $\pm$ \phn\phn8.8 & 5.502 & 7 & 187 & 1.1e-15 & 1.9e-16 & 11 & R20-131 & \nodata & 0 \\
X188 & 13:36:59.649 & -29:51:55.83 & 0.40 & C & n & -12.2 $\pm$ \phn19.4 & -46.9 $\pm$ 16.3 & -45.7 $\pm$ \phn15.3 & 9.152 & 7 & 188 & 1.4e-14 & 2.9e-15 & 3 & R20-140 & \nodata & 0 \\
\hline
\multicolumn{18}{l}{(This table is available in its entirety in machine-readable form.)}
\end{tabular}
\end{center}
\end{table*}


\begin{table*}
\caption{SNR and X-ray Forced Photometry Catalog Column Descriptions\label{tab:forced_columns}}
\begin{center}
\begin{tabular}{lcl}
\hline
\hline
Column Name & Units & Description \\
\hline
Name         &        & Name of SNR \citep{williams19} or X-ray source \citep{long14}.\\
RA           & hh:mm:ss.sss & J2000 RA from external catalog \\
Dec          & ${+}$dd:mm:ss.ss & J2000 Declination from external catalog \\
$\rho$       & kpc    & Galactocentric distance \\
$D$     	 & pc     & Diameter \\
Cl           &        & Visual classification$^a$ \\
Wrn          &        & n = in confused nuclear region \\
F5.5GHz      & \uJy   & Integrated flux at frequency 5.5 GHz, $\Fint$ \\
F5.5Err      & \uJy   & Error on F5.5GHz \\
F9GHz        & \uJy   & Integrated flux at frequency 9.2 GHz, $\Fint$ \\
F9Err        & \uJy   & Error on F9GHz \\
Fint         & \uJy   & Flux at pivot frequency pFreq, $\Fint$ \\
FiErr        & \uJy   & Error on Fint \\
pFreq        & GHz    & Pivot frequency $\nu_0$ where signal-to-noise is maximized \\
nBands       &        & Number of frequency bands with data (1 to 7 of $4\times5.5$, $3\times9$ GHz subbands) \\
$H\alpha(\mathrm{tot})$ & $\mathrm{erg\, cm^{-2} s^{-1}}$ & Integrated $H\alpha$ flux$^b$ \\
$H\alpha(\mathrm{ave})$ & $\mathrm{erg\, cm^{-2} s^{-1} arcsec^{-2}}$ & $H\alpha$ surface brightness \\
Rflag        &        & Radio detection flag$^c$ \\
Radname[1-2] &        & Names of associated radio source(s)$^d$ \\
nXray        &        & Number of associated X-ray sources in \cite{long14} \\
nSNR         &        & Number of associated SNRs in \cite{williams19} \\
\hline
\multicolumn{3}{l}{
\begin{minipage}{0.9\textwidth}
Column descriptions for Tables~\ref{tab:forced_snr} and~\ref{tab:forced_xray}.\\
$^a$ Meaning of visual classification values:
A = object with clear radio counterpart isolated from other sources;
Af = object with faint/marginal but well-aligned radio source isolated from other sources;
B = object at the appropriate position, in a  region that has some emission that most likely arises from other sources;
C = object in confused region of radio emission so that one cannot tell if object is detected;
D = isolated object but no evidence of a radio counterpart.  (Most sources classified as A and the majority of sources classified visually as B will have been detected at 3$\sigma$ or greater.)\\
$^b$ The $H\alpha$ fluxes were obtained from images described by \cite{blair12} and include some contribution
from [N~II]$\lambda\lambda$6548, 6583.\\
$^c$ The detection flag (Rflag) is a bit flag where the bits indicate
whether the association between the radio source and the SNR or X-ray source is unambiguous:
1 = this source may have a match in the radio catalog;
2 = unambiguous match: this source matches only one object from the radio catalog;
8 = mutually good match: this source is best for the other object, and the other object is best for this source.
The most reliable matches will have flag bit 8 set.  All of those will have bit 1 set as well, and most of them
will also have bit 2 set.
See Table~\ref{tab:flags} for more information.\\
$^d$ Multiple radio matches are listed in order of decreasing island overlap.\\
\end{minipage}}
\end{tabular}
\end{center}
\end{table*}

\subsection{Visual Classifications of Sources in Forced Photometry Catalogues}
\label{sec:visual}

\begin{figure*}
\centering
\includegraphics[width=1\textwidth]{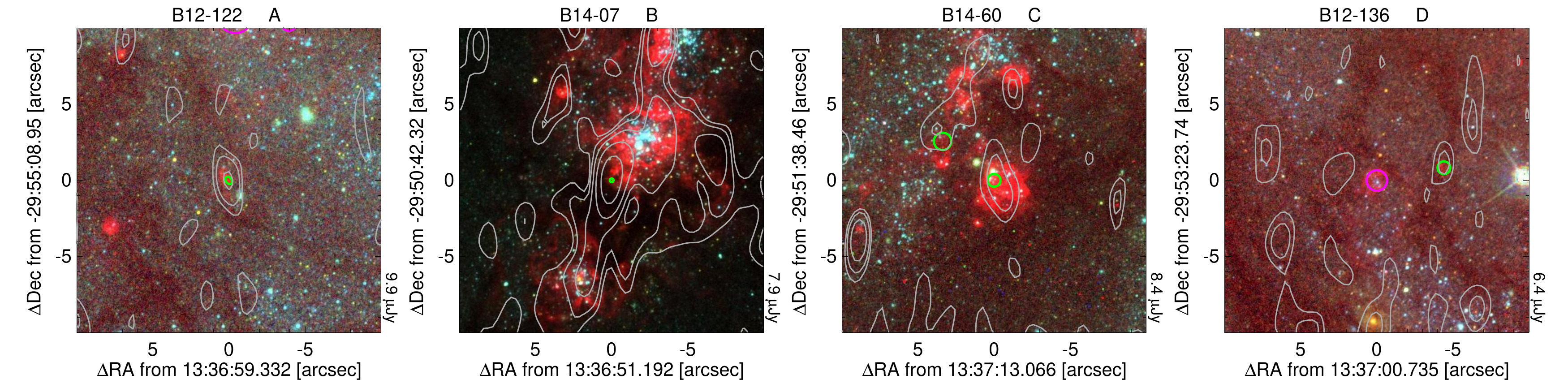}
\caption{This four panel figure shows examples of the A, B, C, D classification groups from our visual inspection of radio emission relative to optical SNR positions. The title gives the SNR name and visual classification.  In these examples, the background images are all from {\it HST}/WFC3, with \ha\ (prior to continuum subtraction) in red, V-band in green, and B-band in blue.  (Hence, star colors are approximately correct.) Gray radio contours increase by factors of two, with the lowest contour value given on the right side of the box. Regions are shown for the sizes and positions of the known optical SNRs. The objects graded as `A' are isolated SNRs with solid radio detections while `D' grades indicate isolated SNRs that were not detected.  Intermediate cases include `B' where the SNR is detected but somewhat confused by adjacent radio emission, while objects in `C' class are too confused to allow characterization of the SNR radio emission.}
\label{fig:class_examples}
\end{figure*}

Because of the complexity of the radio emission seen, there are bound to be places where the source finding algorithms and the derived fluxes are compromised.  Since it is desirable to know which sources are clean, isolated detections and which are not, we decided to perform a visual classification.  Furthermore, by comparing against lists of known optical sources like the optical SNR catalogue \citep{williams19} or  the {\em Chandra} X-ray source catalogue \citep{long14}, we can clearly distinguish which sources are non-detections in the radio and thus provide hard upper limits. We can also treat any detected emission at the positions of sources in complicated regions as upper limits as well.

We performed this visual classification by projecting the radio contours onto {\it HST}/WFC3 (or Magellan if outside the {\it HST} footprint) image data, color-coded to show \ha\ emission in red, V-band in green, and B-band in blue. The \ha\ frame prior to continuum subtraction was used so that star colors would be somewhat representative. DS9 regions for a) the optical SNR list, and b) the X-ray source list, were displayed to guide the visual inspection and help judge the quality of the alignment of any radio emission with the SNR or X-ray source.  We adopted a simple A, Af, B, C, D grading system for the quality of the radio association with each object such that A indicated an isolated source with clear radio counterpart, Af indicated a faint radio counterpart, B indicated clear radio emission from the source but with some confusion from nearby emission, C implied the source is in such a confused or complex region of extended radio emission that the veracity of a radio detection could not be established from visual inspection alone, and D indicated isolated sources with no detected radio emission.

Figure~\ref{fig:class_examples} shows an example of each of these classifications for several SNRs to provide a visual context.  With this classification, we can inspect the properties of the different groupings separately in what follows.

\section{Results and Analysis}
\label{sec:results}

As noted earlier, radio emission in M83 is concentrated along the spiral arms (cf. Figure~\ref{fig_overview}).  The two main sources of emission along these arms should  arise from either \hii\ regions or SNRs.  There are also sources that appear to be between the main spiral arms.  Some of these could be extragalactic background sources, but even here, many can be associated with \hii\ regions in spurs that protrude from the main spiral arms and hence, are likely to be intrinsic to M83.

\subsection{Comparison With Previous Radio Studies of M83}
\label{sec:previous_radio}

Prior to this work, \citet{maddox06} presented a catalogue of radio sources from M83, using 1.45 and 4.9~GHz VLA observations taken between 1981 and 1998. Here, we compare their catalogue and implied spectral indices to our results. Our radio observations probe lower luminosities and higher angular resolution, so we detect more radio sources both due to an increase in sensitivity, and also due to our ability to resolve some of the emission regions into multiple smaller sources. 
To aid comparison, we convolved our image with a Gaussian the same shape and size as the VLA beam; qualitatively their image and our convolved images look very similar.
Of the 55 sources identified by \citet{maddox06}, 49 lie within 2\arcsec\ of sources in our catalogue (once an allowance of about $0\farcs5$ is made for a systematic offset in the absolute positions). 
The \textit{AltName} column in our radio catalogue indicates sources that match a \citet{maddox06} object (Table~\ref{tab:radio_catalogue}).
Our flux densities and those measured by \citet{maddox06} appear correlated, although the latter work reported higher flux densities for fainter radio sources and lower flux densities for brighter sources than we do, which is likely due to the difference in surface brightness sensitivity. (Malmquist bias may also play a role for objects near their detection limit.)  Note that \citet{maddox06} published only peak pixel flux densities rather than integrated flux densities, which means their values will be systematically lower than our catalogue flux densities for extended sources.

The six \citet{maddox06} sources that do not have matches within 2\arcsec\ in our catalogue are M-5, M-21, M-30, M-32, M-37, and M-44.  Some of these sources are certainly variable.  M-5 is the historical SN 1983N (see section \ref{sec:histsn}), which was  detected only in 1983/84 and faded below the detection limit in our map.  M-32 is in the nuclear region, so the radio emission is confused, although there is plenty of radio flux nearby.  It is $2^{\farcs}9$ from our nearest catalogue source.  M-21, M-30, and M-37 are in regions with evidence of radio emission but are not close to peaks in our maps.  M-44 shows no evidence of any radio emission in our ATCA data (and also lies in a very dusty region with little optical emission).

\begin{figure}
	\centering
\includegraphics[width=0.95\columnwidth]{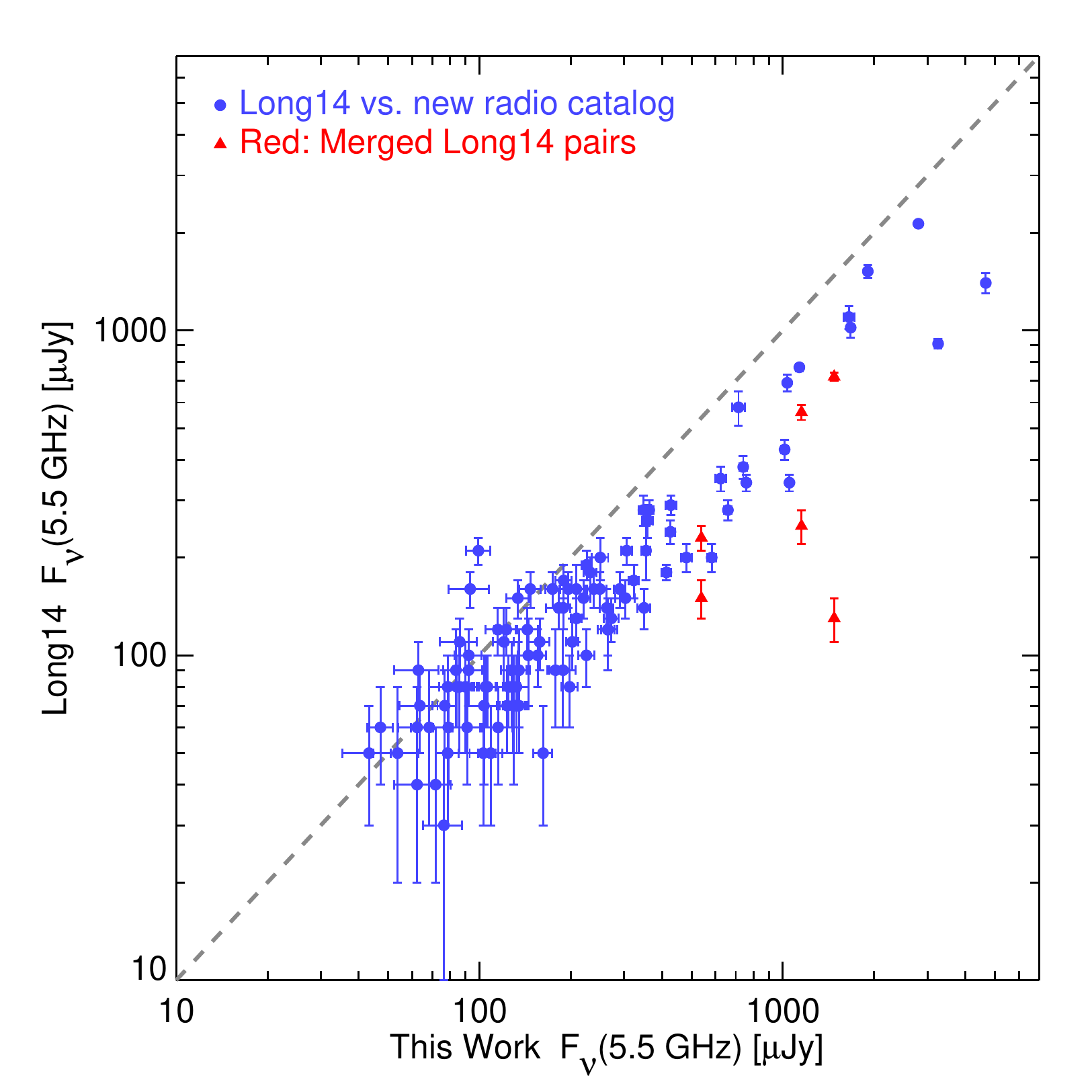}
\caption{A comparison between ATCA fluxes at 5.5~GHz reported by \citet{long14} and those reported in the current catalogue. The red triangles indicate three object pairs from \citet{long14} that are merged into a single source in our new catalogue.
\label{fig_L14_compare}
}
\end{figure}

Of the 109 radio sources identified in our earlier ATCA survey from 2011 \citep{long14}, all but one have positions that lie within one of the islands for a detected source in our new catalogue.  The single missing source, A048, was reported as having a flux density at 5.5~GHz of $0.03\pm0.02$ mJy in \citet{long14}, and is probably not a true detection.  The median offset between the positions in the catalogues is 0\arcsec.3.  For three objects in our new catalogue, two sources from the older catalogue are merged into a single new source: R20-069 (A026$+$A029), R20-155 (A063$+$A064), and R20-181 (A077$+$A079).  As shown in Figure~\ref{fig_L14_compare}, the flux densities are fairly well correlated except that there is a tendency for sources to appear brighter in the new catalogue. This difference is most likely due to this new work including observations taken with multiple telescope configurations providing better UV-plane coverage; in particular, our 1.5D configuration observations provided shorter baselines.  The cases where the new catalogue merges sources from the older catalogue are also likely examples where more extended emission bridged the gap between sources that appeared separate in the earlier data.

\subsection{Supernova Remnants in the Survey}

As one might expect for a nearby, relatively-face on spiral galaxy with a star formation rate of 3-4 \MSOL yr$^{-1}$  that has hosted six or seven SNe in the last century, M83 contains a large number of SNRs, most of which have been identified optically as nebulae that have high ratios of \sii:\ha\  emission. 
 Our current list of published optical SNRs and SNR candidates totals 304 emission nebulae \citep{blair12,blair14,dopita10}.\footnote{For the purposes of this study we consider only objects identified as SNRs on the basis of elevated \sii:\ha\ ratios.  In an attempt to identify very young SNRs dominated by emission from SN ejecta, \citet{blair12} also created a candidate list of SNRs based on elevated \oiii:\hb\ ratios; a handful of these objects has turned out to be normal SNRs based on follow-up spectroscopy and these are included, but other objects from that list are excluded.  One exceptional object,  the very young SNR B12-174a \citep{blair15} stood out from both \sii:\ha\ and \oiii:\hb\ criteria \citep[see][]{winkler17}, and is included.  The SNR list used here is essentially the same as described by \citet{williams19}, except that we include two objects identified as SNRs by \citet{dopita10} that were inadvertently omitted: D10-N-03 and D10-N-04.}  
The increase in numbers arises primarily from the addition of a large population of small-diameter and \feii-identified SNR candidates found by analyzing images obtained with HST \citep{blair14}.  

SNRs  are the brightest radio sources in our Galaxy, and so many SNRs are expected to be present within our radio catalogue of M83.   In our earlier work based only on the 2011 ATCA data, we identified 23 radio sources which we suggested were SNRs, based on their location within 2\arcsec\ of  one of the 225 optical SNR candidates identified at that time  using ground-based data \citep{blair12}.  Most of the radio sources were also associated with X-ray sources, which we argued supported the identification of the radio sources as SNRs.   Of the optical SNRs, 117 (of 118) have been confirmed in the sense that spectra have been obtained that verify estimates of the \sii:\ha\ ratio from imaging observations \citep{winkler17}.   

We have searched for SNRs in our radio images in two ways:  First, we have looked for positional coincidences with the optical SNR catalogue, and find there are 83  sources in the general radio catalogue that are spatially coincident with optical SNRs; of these, 11 are in the complex nuclear region of the galaxy (and hence could be chance coincidences) and 72 are outside the nuclear region.

Secondly, we have used the forced-photometry approach, where we have extracted the radio flux density at the positions of known SNRs (see Sec.~\ref{sec:forced_photometry}).  Taken at face value, 125 of these sources are detected at the 3$\sigma$ level.  Unfortunately, this number overestimates the number of detected SNRs, because SNRs are frequently located in or near regions of substantial thermal emission from \hii\ regions.  Eliminating the SNRs in these ``confused'' regions of the radio data,
for the 199 isolated sources in the sample (classes A, B, and D only --- see section~\ref{sec:visual}), we detected 64 of these as radio sources at the 3$\sigma$ level or higher. In M33, \citet{white19} were able to isolate many of the SNRs from \hii\ regions based on the radio spectral index (as well as the  better spatial scale for the much closer M33) but as discussed earlier we have not been able to do this in M83. 

\begin{figure}
	\centering
\includegraphics[width=0.9\columnwidth]{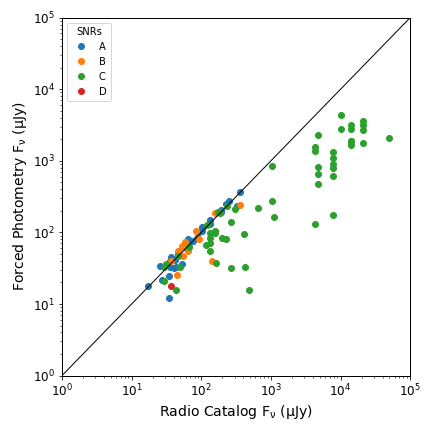}
\includegraphics[width=0.9\columnwidth]{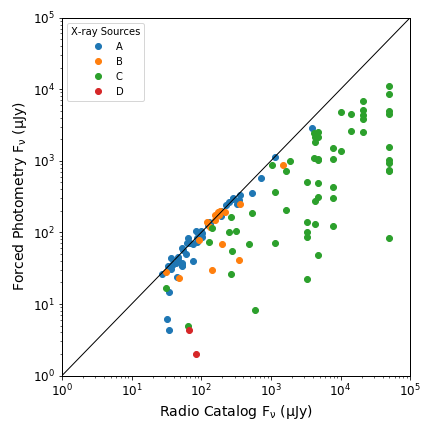}
\caption{A comparison between the forced photometry and radio catalogue fluxes for SNRs (top panel) and for X-ray sources (lower panel).  Only sources that have have apparent detections of at least  3-$\sigma$  in the relevant forced photometry catalogue are plotted.  The color of each point indicates the visual classification of the object. \label{fig_radio_vs_forced}}
\end{figure}

The top panel of Figure \ref{fig_radio_vs_forced} shows a comparison of the SNR radio fluxes from the general catalogue to those from the forced photometry catalogue.   As one might expect,  the derived fluxes of 199 isolated sources (A, B, and D classes) are similar in the two versions of the catalogue, but the 105 confused sources (class C) have higher fluxes in the general catalogue than in the forced photometry one.  We take that to mean that the forced photometry provides the best actual estimates of radio emission from the  SNRs in M83, and confirms that all of the values for SNRs labeled as confused (class C) should be taken as upper limits. 

Of the SNRs in the forced photometry catalogue, there are eleven that were discussed in detail by \citet{soria20}.  These are, in right ascension order, B12-045, B12-096, B12-098, B12-122, B12-124, D10-N-16, B12-143, B12-146, B14-45, B12-169, and B12-209.  These were selected because of unusual optical morphologies that suggest they could be microquasars masquerading as SNRs.  One of them, D10-N-16 (a.k.a. MQ1), is a definite microquasar discussed by \citet{soria14}.  Another, labeled `S2' by \citet{soria20}, is a combination of two optical SNR candidates, B12-096 and B12-098, which are probably two lobes of emission from a single central object, most likely also a microquasar.  The remaining objects were considered more likely to be conventional SNRs, albeit with odd morphology in the optical.  We have included all of these objects here in our list of SNR candidates.

\subsection{X-ray Sources in the Survey}

As noted previously, there were 424 X-ray sources in the deep Chandra survey of M83 \citep{long14}.  Based on our visual inspection of the radio data, we classified 350 of these sources as relatively isolated  (A, B, or D) and 74 as lying in confused regions of the radio image where forced photometry was likely to be unreliable (class C).  The forced photometry estimates indicated 129 of the X-ray sources were detected at greater than $3\sigma$ in the radio image, of which 75  were in unconfused regions.  All of these 75 had been classified as detected (A) or likely detected (B); none of the X-ray sources classified from visual inspection as undetected exceeded the 3-$\sigma$ threshold for radio detection.

\begin{figure}
	\centering
\includegraphics[width=0.98\columnwidth]{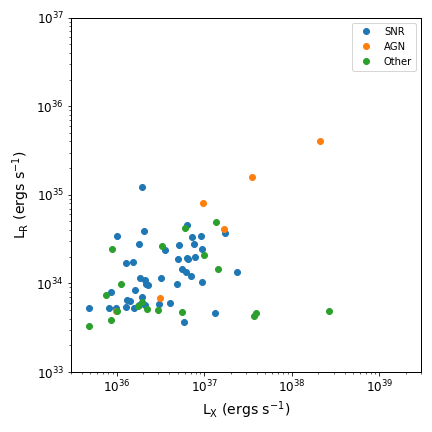}
\caption{A comparison of X-ray and radio luminosities for isolated sources.  Only sources detected at 3$\sigma$ or greater at 5.5 GHz are shown
\label{fig_xray_lum}}
\end{figure}

The X-ray and radio luminosities of the isolated sources that were detected at radio wavelengths is shown in Figure \ref{fig_xray_lum}. Of the 75 sources, 50 were classified as SNRs and 6 as either a background AGN or galaxy by \citet{long14}.   One source, X321, was classified as an X-ray binary by \citet{long14} due to its variable brightness and hard X-ray spectrum.  But X321 is also coincident with the SNR B12-179, a small, compact optical SNR on the outer fringes of bright \hii\ emission; therefore, it is possible that the radio emission actually arises from the SNR since, with some exceptions, X-ray binaries are generally not strong radio sources.  Of the 18 remaining sources detected in unconfused regions of the map, it is interesting to note that the soft source X124 lies within 1\arcsec\ of the SNR B12-079 and X219 lies close to the SNR B14-031, although only X124 has a soft X-ray spectrum.  In any event, what is clear from this discussion is that the X-ray sources in M83 that are also associated with radio sources are almost exclusively SNRs.

\begin{figure}
	\centering
\includegraphics[width=0.9\columnwidth]{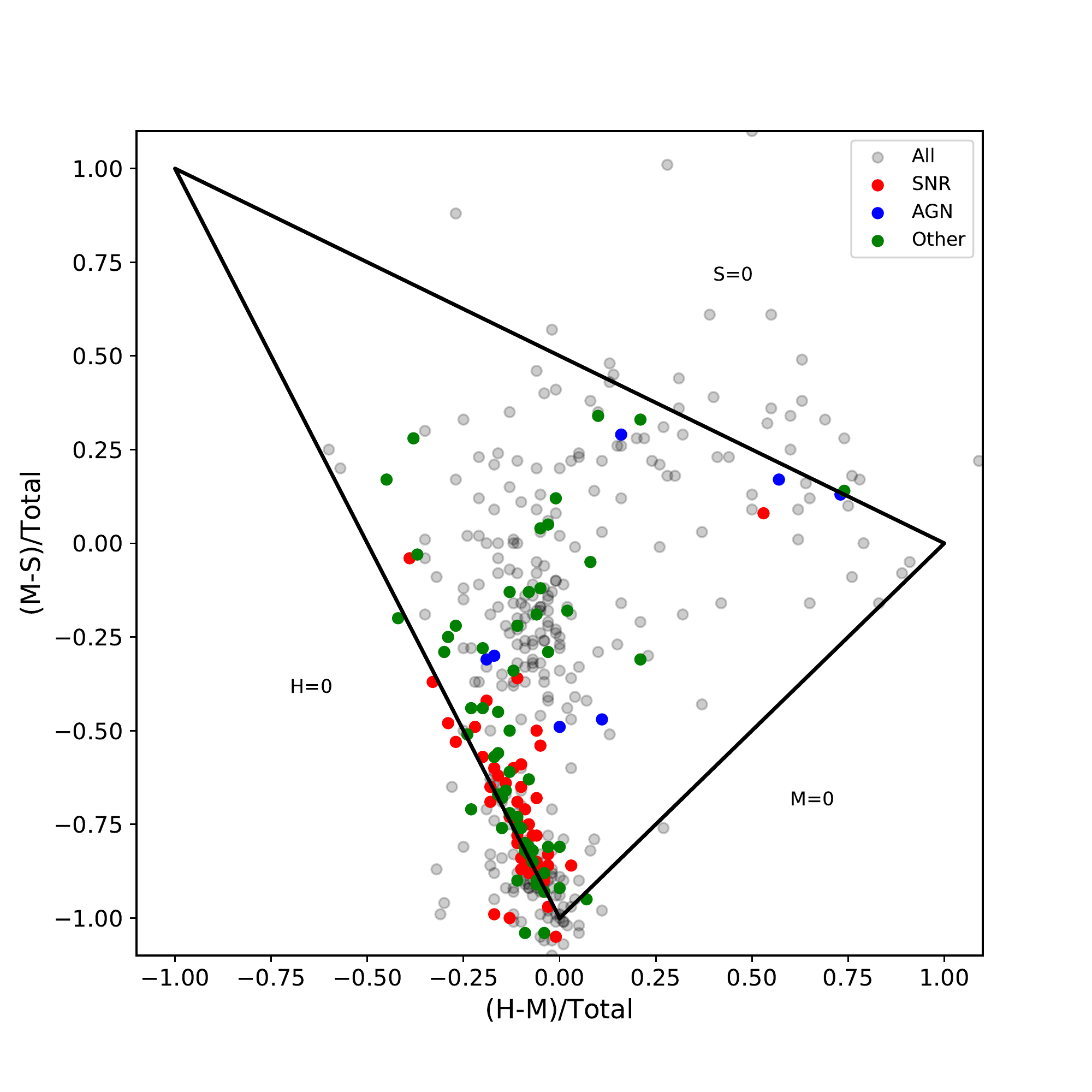}
\caption{Hardness ratios for X-ray sources in M83 as determined by  \citet{long14},  comparing the X-ray flux in soft (S = 0.35 - 1.1 keV), medium (M = 1.1 - 2.6 keV), hard (H = 2.6 - 8 keV) and total (T = S + M + H) energy bands.  Here, X-ray sources that are detected as radio sources are shown in red if they are known SNRs, blue if they are AGN, and green otherwise.  The remaining X-ray sources (in grey) have no detected radio counterparts.
\label{fig_xray_color}
}
\end{figure}

An X-ray hardness diagram, comparing the X-ray flux in soft (S = 0.35 - 1.1 keV), medium (M = 1.1 - 2.6 keV), hard (H = 2.6 - 8 keV) and total (T = S + M + H) energy bands is shown in Figure~\ref{fig_xray_color}.  Most of the sources with $(M-S) > -0.5$ are thought to be X-ray binaries or background AGN, while most of the sources that have $(M-S)/T \le- 0.5$ are thought to be SNRs.   Objects that are detected in the radio (whether or not they are in confused regions) and that were identified as SNRs by \citet{long14}  are shown in red; they cluster, as expected, in the region of the diagram containing soft X-ray sources.  Objects detected at radio wavelengths identified by \citet{long14} as AGN are shown in blue, and X-ray point sources that we detect but which are unclassified are plotted in green.  Of the 64 unclassified sources that are detected at 3$\sigma$, 37 are in the region expected to be SNRs.  Of these, there are eight that are isolated.  These are the strongest candidates for being previously unrecognized SNRs.\footnote{The objects are X020, X066, X074, X154, X164, X177, X179,  and X282.}

\subsection{Background Sources in the Survey}

Most of the radio emission in our radio image of M83 arises from sources that belong to the galaxy itself.  Many sources clearly follow the spiral arm structures or fall in regions of dust or star formation.  However, clearly some background sources would be expected from a survey of this sensitivity.  The most obvious example is R20-119, seen to the NW of the nucleus, which is the core of a background FRII radio galaxy (as discussed in \citealt{long14}), along with three other sources, R20-111, R20-115, R20-124, that are associated with the lobes of the same radio galaxy.\footnote{R20-115 has a SNR association also, but this association is spurious.} Four other sources, R20-024, R20-263, R20-264, and R20-265, located at the outskirts of our field of view are easily recognized as background galaxies in optical images.

To estimate the total number of background sources expected, we integrated the semi-empirical differential flux density distribution of \citet{wilman08} using the $4\sigma$ detection limit versus radius shown in Figure~\ref{fig:noise_vs_dist} (similar results are obtained from the source number distributions of \citealt{condon12} and \citealt{condon84}). We interpolated between the source number distributions at 4.8~GHz and 18~GHz from \citet{wilman08} to obtain the source number distribution at our effective central frequency as a function of radius (also shown in Figure~\ref{fig:noise_vs_dist}). Based on this, we expect a total of 34 background sources over the entire image area, with about 22 sources in the region within $\sim 4\farcm 5$ of the galaxy center where most of our catalogue sources are found. Thus the contamination of the sample by background AGN is modest ($<10\%$) except at the outer edges of the galaxy.

In addition to the obvious background sources noted above, we performed a visual search of the general radio catalogue and noted any sources that did not have obvious optical counterparts (excluding radio sources projected onto obvious galaxy structures like dust lanes).  There were 32 such sources in addition to the obvious sources mentioned above, which is totally consistent with expectations from above.

This situation is very different than in M33, as discussed by \citet{white19} in their analysis of a VLA survey of that galaxy.  In M83, most of the radio sources that are detected as X-ray sources are likely SNRs, whereas in M33 the majority of the radio sources detected as X-ray sources are extragalactic interlopers behind the galaxy.  In M33, the fraction of AGN was $\approx 30\%$ at an apparent luminosity in M33 of $\sim 10^{36}$ erg s$^{-1}$, and $\approx 10\%$ at an apparent luminosity of $\sim 10^{38}$ erg s$^{-1}$ \citep{long14}. This difference, of course, is understandable from the difference in angular sizes of the two galaxies.

\subsection{The Nuclear Region}
\label{sec:nucleus}

\begin{figure*}
\centering
\includegraphics[width=0.35\textwidth]{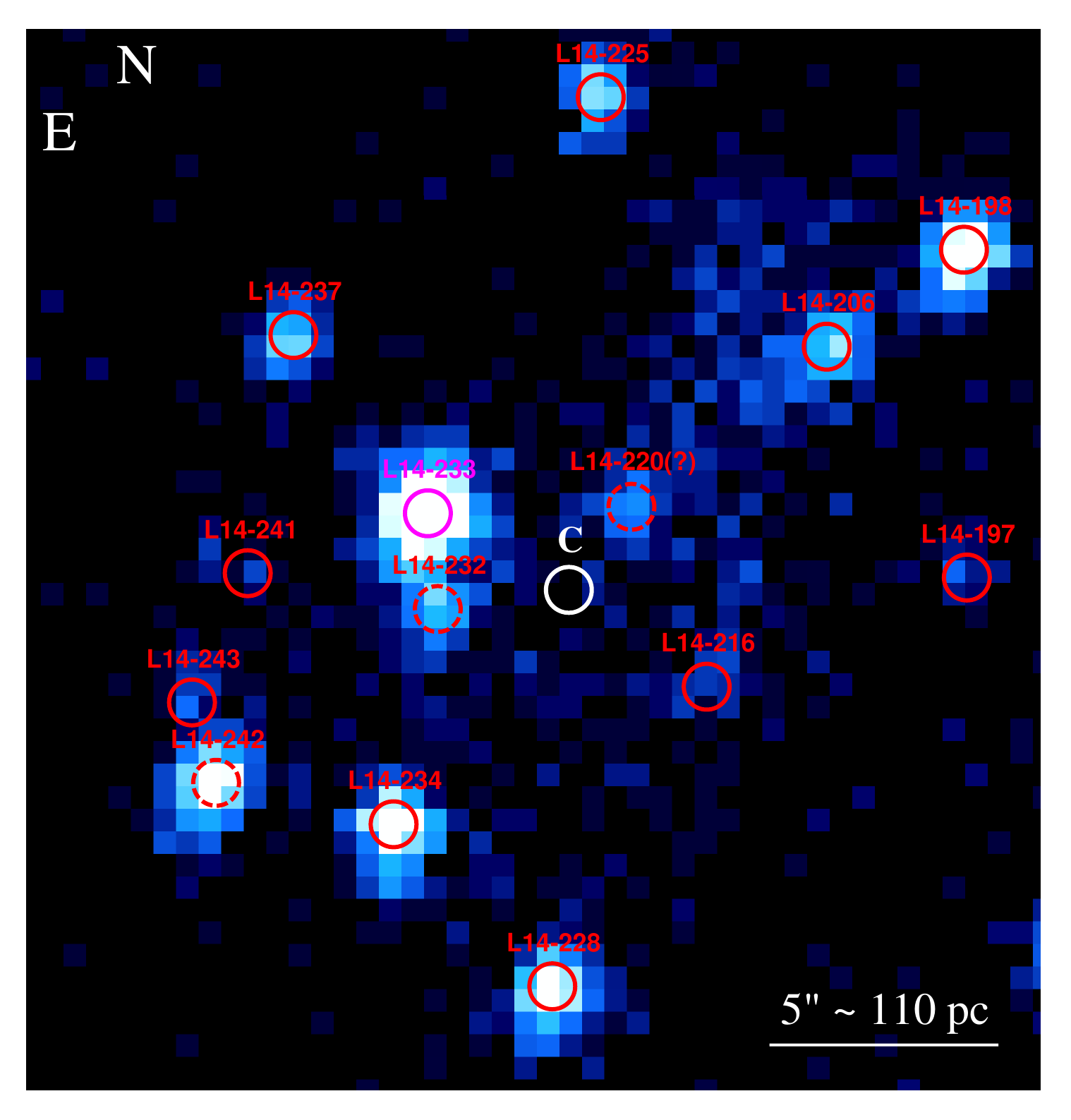}
\hspace{-0.5cm}
\includegraphics[width=0.35\textwidth]{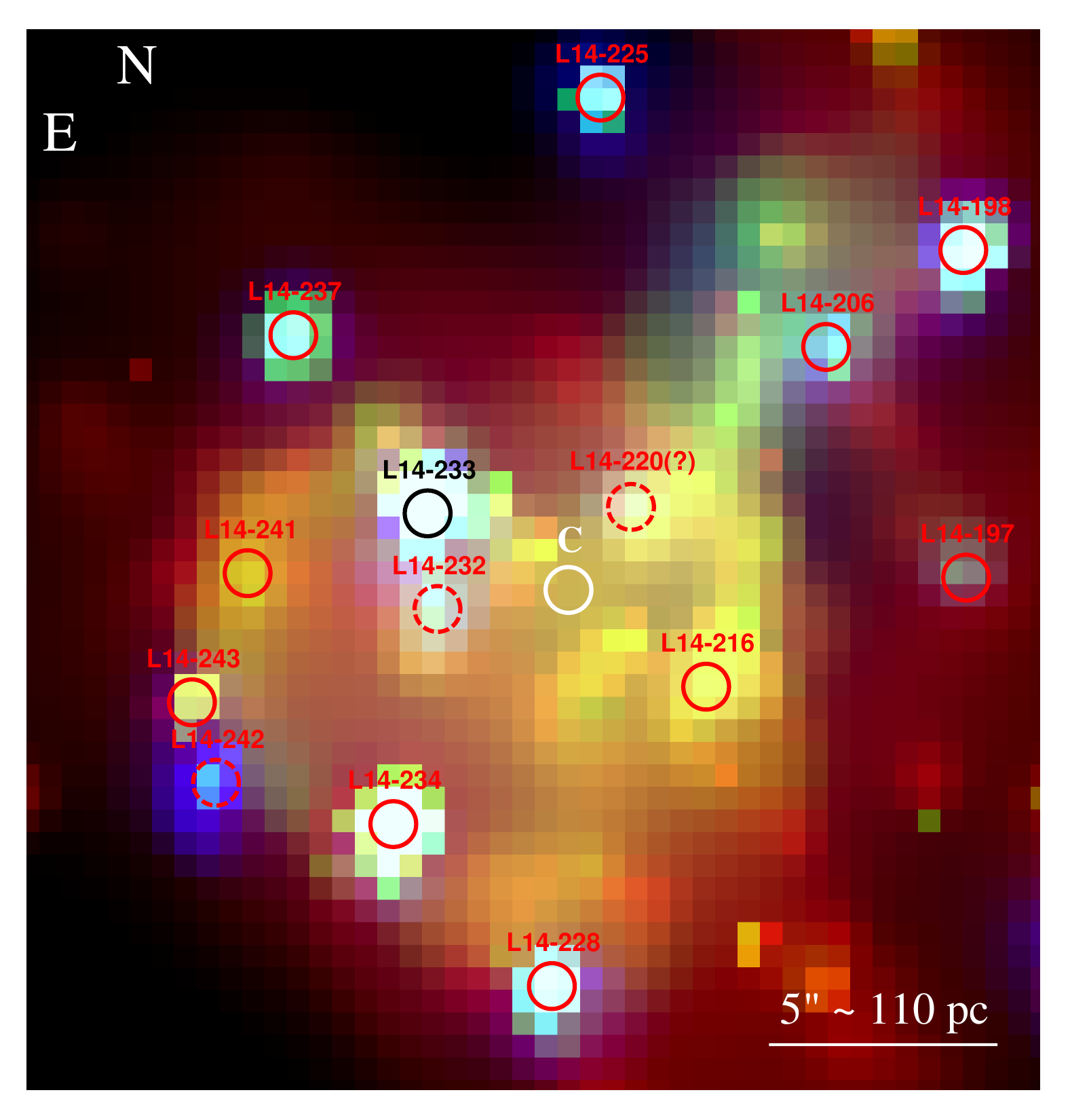}\\
\vspace{-0.3cm}
\includegraphics[width=0.35\textwidth]{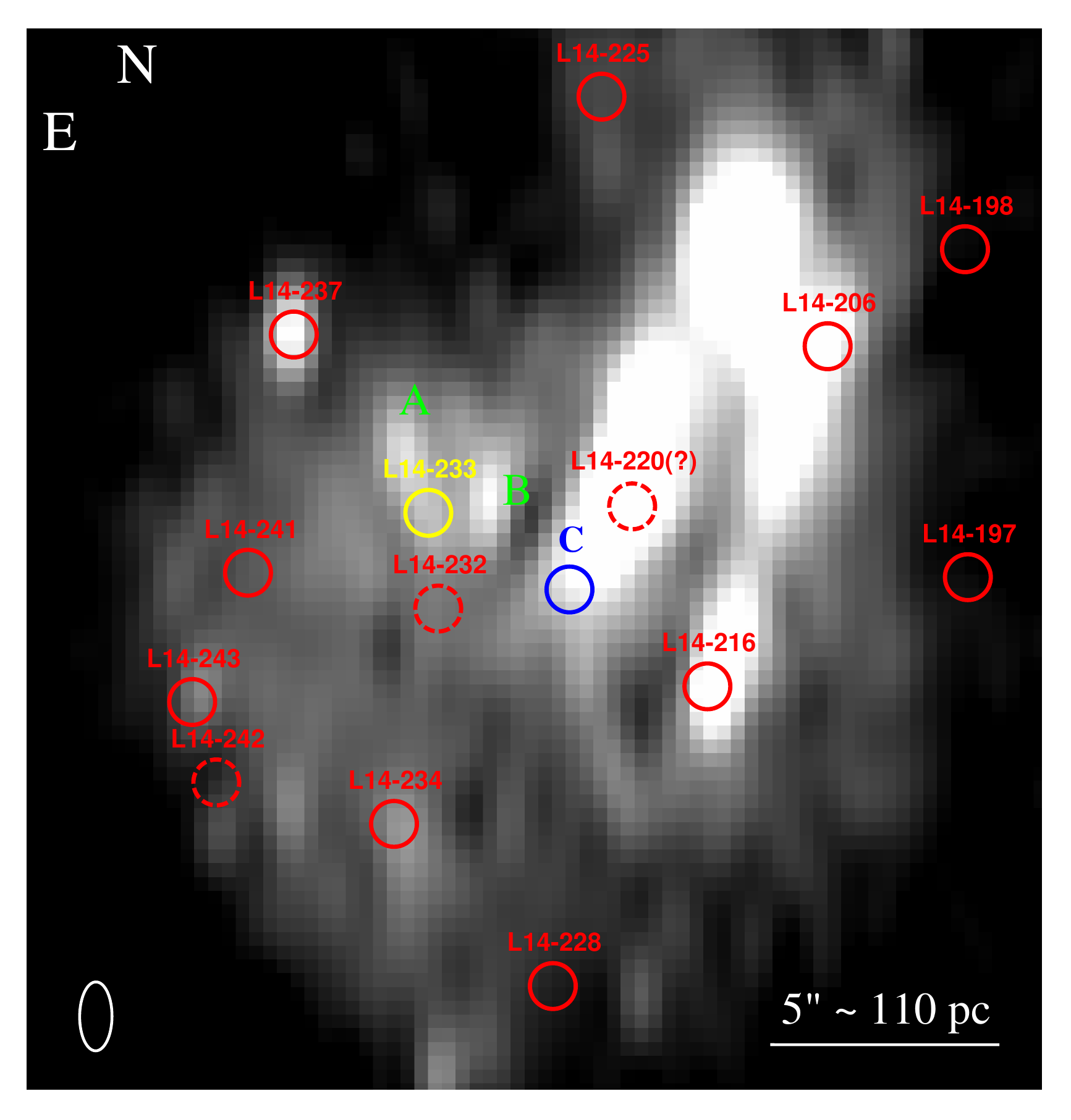}
\hspace{-0.5cm}
\includegraphics[width=0.35\textwidth]{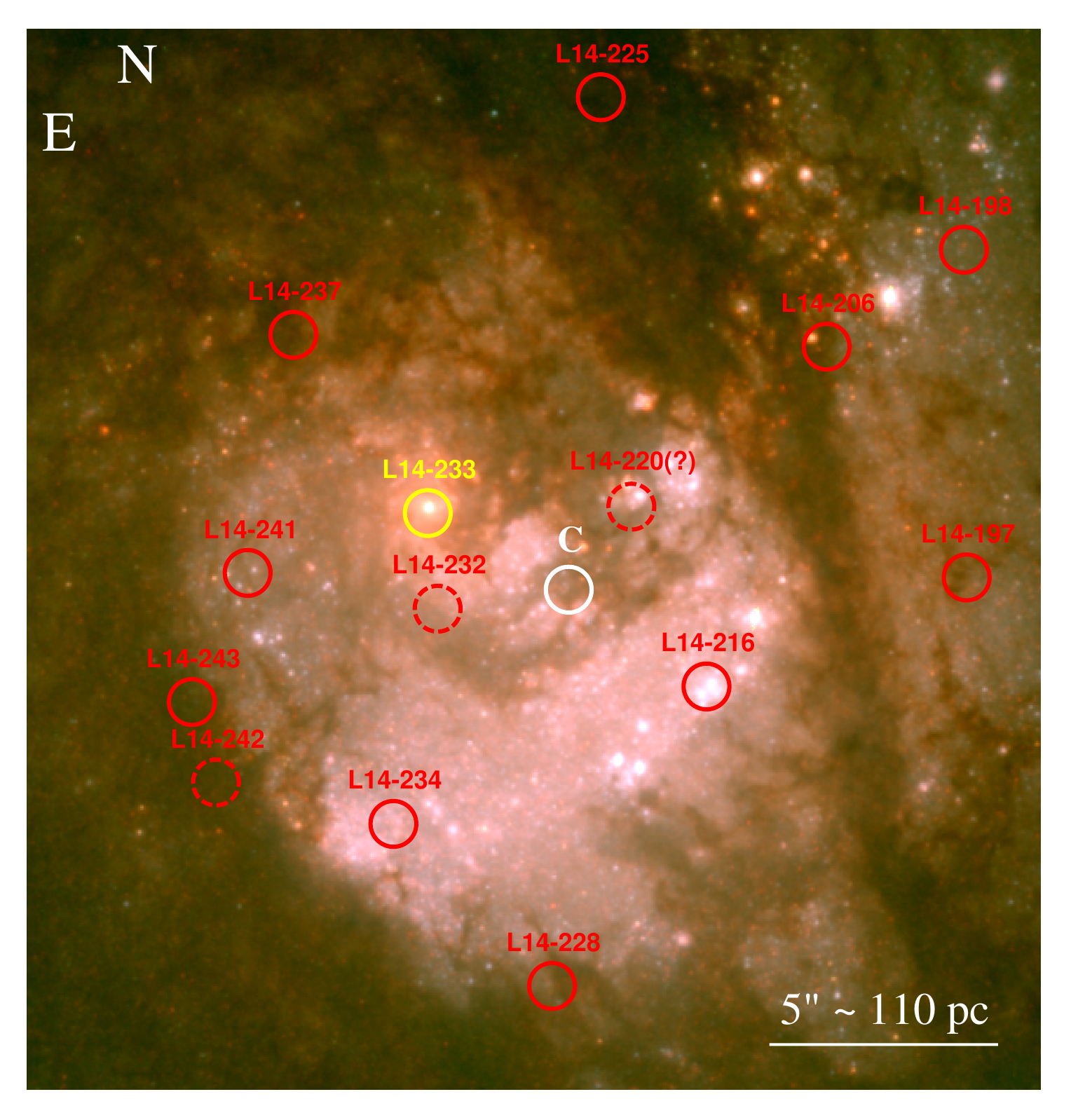}\\
\vspace{-0.3cm}
\hspace{0.01cm}
\includegraphics[width=0.35\textwidth]{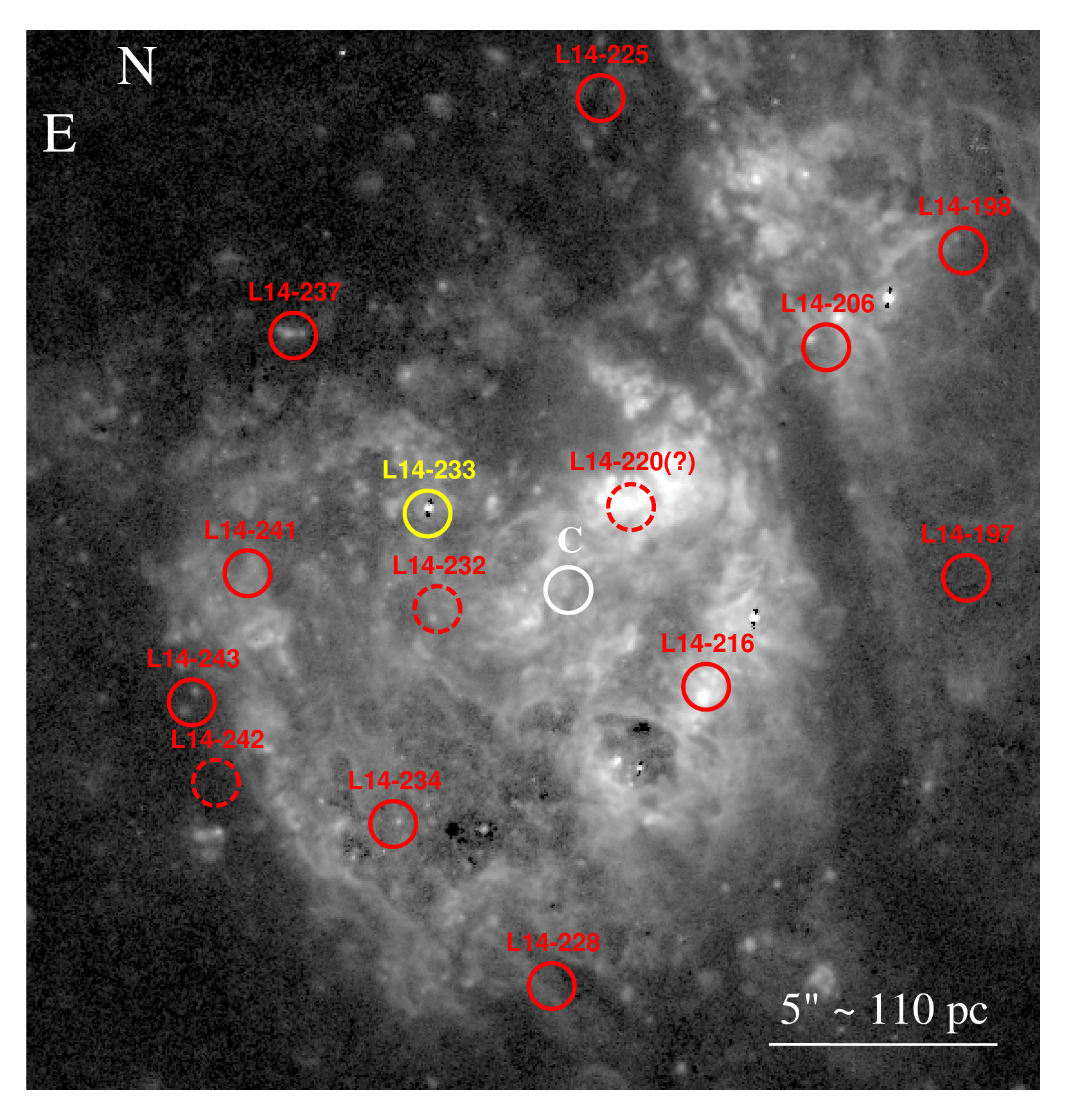}
\hspace{-0.5cm}
\includegraphics[width=0.35\textwidth]{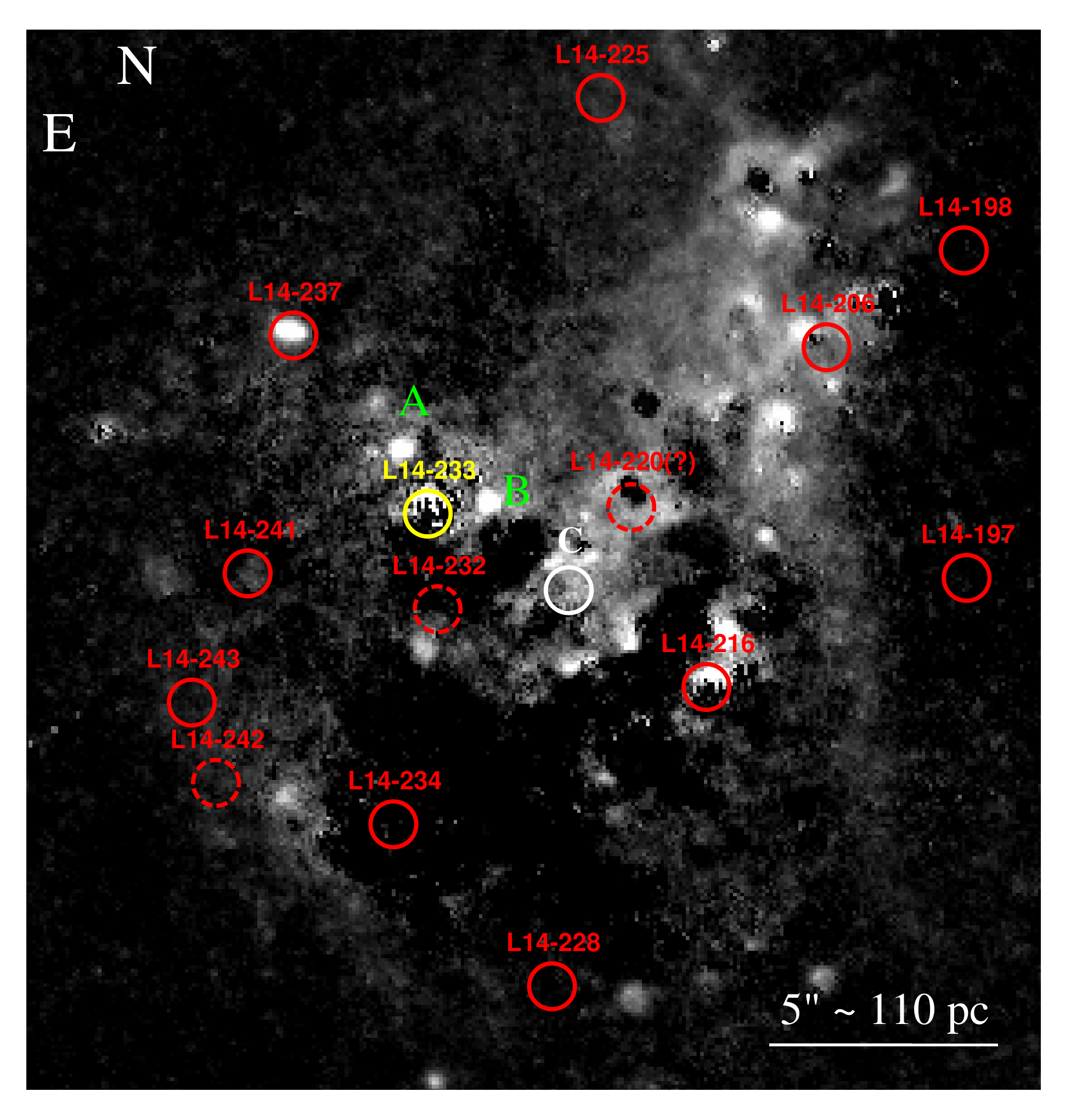}
\caption{
A 45\arcsec\ region of M83 centered on the nuclear region.
{\bf Top left:} stacked {\it Chandra} image in the 2.6--8 keV band. The
circles and labels correspond to \protect\citet{long14} sources (but omitting
sources not detected in the hard band); the diameter of each circle
is $1\farcs0$. L14-233 corresponds to the optical nucleus; L14-237
to the microquasar MQ1 \protect\citet{soria14}. Dashed circles correspond
to sources whose positions (derived from the hard band only) appear
slightly different ($0\farcs4-0\farcs8$) from what was reported in
the L14 catalogue (based on the full band). The white circle labelled
C is the location of the photometric center \protect\citep{knapen10}. {\bf Top
right:} adaptively smoothed {\it Chandra} image, with red = 0.35--1.1
keV, green = 1.1--2.6 keV, blue = 2.6--8 keV; circles and labels
are the same as in the top left panel. {\bf Left middle:} ATCA 9 GHz flux
density map; the synthesized beam is overplotted on the bottom left
of this panel. The unresolved sources A and B near the nucleus
correspond to two candidate SNRs identified by \protect\citet{piqueras12} and discussed in Section 4.5.
{\bf Right middle:} {\it HST} broad-band image; red = F814W, green =
F555W, blue = F438W. {\bf Bottom left:} continuum-subtracted {\it HST}
image in the F657N filter. {\bf Bottom right:} continuum-subtracted {\it
HST} image in the F164N filter; notice the prominent \feii\ emission
from the two candidate SNRs A and B.
\label{fig_nucleus_6pan}}
\end{figure*}

The nuclear starburst region of M83 is extremely complex, even as viewed at {\it HST} spatial resolution \citep{dopita10}, so it is not surprising that our ATCA data do not totally resolve the radio structure. In Figure~\ref{fig_nucleus_6pan}, we show a six-panel multiwavelength view of a 40\arcsec\ field centered on the nucleus that demonstrates the complexity. Appendix A provides details about the possible mass and position of a purported supermassive BH, which aligns with source L14-233 in the figure.  Interestingly, the kinematic and photometric center is offset to the SW from this position, at the position marked `C' in the Figure.     

The brightest radio enhancement, which is somewhat extended, seems to align with a region of heavy extinction to the upper right of field center.  Lower level diffuse radio emission seems to fill the same cavity seen in the optical continuum images.  One source that clearly aligns with a known source is indicated by the red circle marked L14-237 in the Figure, which is the microquasar MQ1 on the NE fringes of the nuclear region, that was identified with {\it HST} and has both optical and X-ray counterparts in addition to being a radio source \citep{soria14}.

Two other radio sources with a similar 9 GHz flux density of $\approx0.30$ mJy are located $\sim1\farcs5$ to the NNE and $\sim1\farcs5$ to the WNW of the optical nucleus, respectively; they are labelled as A and B in Figure~\ref{fig_nucleus_6pan}. The same sources are clearly visible as point-like sources (diameter $\lesssim 7$ pc) in the continuum-subtracted {\it HST}/WFC3 F164N \feii$\lambda1.644 \mu$m panel of the Figure\footnote{We determine the following coordinates from both the ATCA and {\it HST} images: for source A, RA $=$ 13$^{h}$37$^{m}$00$^{s}$.93, Dec $= -29^{\circ}51^{\prime}54\farcs6$; for source B, RA $=$ 13$^{h}$37$^{m}$00$^{s}$.78, Dec $= -29^{\circ}51^{\prime}55\farcs8$.}. They correspond to the \citet{piqueras12} sources labelled E and F, observed with the SINFONI integral field spectrograph on the Very Large Telescope.  \citet{piqueras12} suggested these two sources were SNRs, which is supported by our detection of them as radio and strong \feii\ emitters. We measured continuum-subtracted fluxes in the \feii$\lambda1.644 \mu$m line of $(3.1 \pm 0.6) \times 10^{-15}$\,erg\,cm$^{-2}$\,s$^{-1}$ and $(3.9 \pm 0.6) \times 10^{-15}$\,erg\,cm$^{-2}$\,s$^{-1}$, for sources A and B, respectively. Thus, the two sources are brighter in \feii\ than all of the young SNRs listed in Tables 3--4 of \citet{blair14}\footnote{Although our SNR catalogue in M83 includes the nuclear SNRs identified by \citet{dopita10} from {\it HST}/WFC3 data, a full assessment of the nuclear region including the \feii\ imaging data has yet to be done.  Hence, these two SNR identifications, as well as other possible \feii\ SNR candidates in this complex region, are not yet listed in our SNR catalogue.}  The only individual source in M83 slightly brighter in \feii\ than those two young SNRs is the microquasar MQ1.

A peak in the radio emission is seen in the star-forming arc $\sim 2\farcs5$ to the north and $\sim 1\farcs5$ to the west of the photometric/kinematic nucleus (C). It is approximately consistent with the X-ray source L14-220 \citep{long14}, and coincides with the peak of the Br$\gamma$ emission labelled ``Aperture A'' in \citet{piqueras12} (also visible in the H$\alpha$ panel of  Figure~\ref{fig_nucleus_6pan}). We associate this enhanced emission with a group of a half dozen massive star clusters at that location, with a characteristic age of $\sim $5--6 Myr and masses up to $\sim 10^5 M_{\odot}$ \citep{harris01}.

Another knot of strong radio emission is located $\sim 2\farcs3$ to the south and $\sim3\farcs0$ to the west of the photometric/kinematic nuclear position, corresponding to the X-ray source L14-216. We estimate a peak pixel 9 GHz flux density of $\sim 1.2$\,\mJybeam\ above the local background.  From inspection of the {\it HST} images, we find that it corresponds to another group of young star clusters in the outer starburst ring, with a characteristic age of $\sim $2 -- 5 Myr and masses up to $\sim 9 ~ \times~ 10^4 M_{\odot}$ \citep{wofford11}.

While additional knots of \feii\ emission (likely SNRs) lie within the large radio enhancement that dominates the upper right quadrant of the radio panel in Figure~\ref{fig_nucleus_6pan}, it is not clear how much of this emission is due to possible SNRs and how much may correspond to \hii\ region emission that is being blocked by the prominent dust lane in this region.  With the large number of optical SNRs identified in the nuclear starburst region by \citet{dopita10} and the complex X-ray diffuse and point source population that includes likely SNRs and X-ray binaries \citep{long14}, this region is worthy of a more detailed study that is beyond the scope of this paper.

\subsection{The Remaining Sources: H~II Regions and Unidentified SNRs}

\begin{figure*}
\centering
\includegraphics[width=0.495\textwidth]{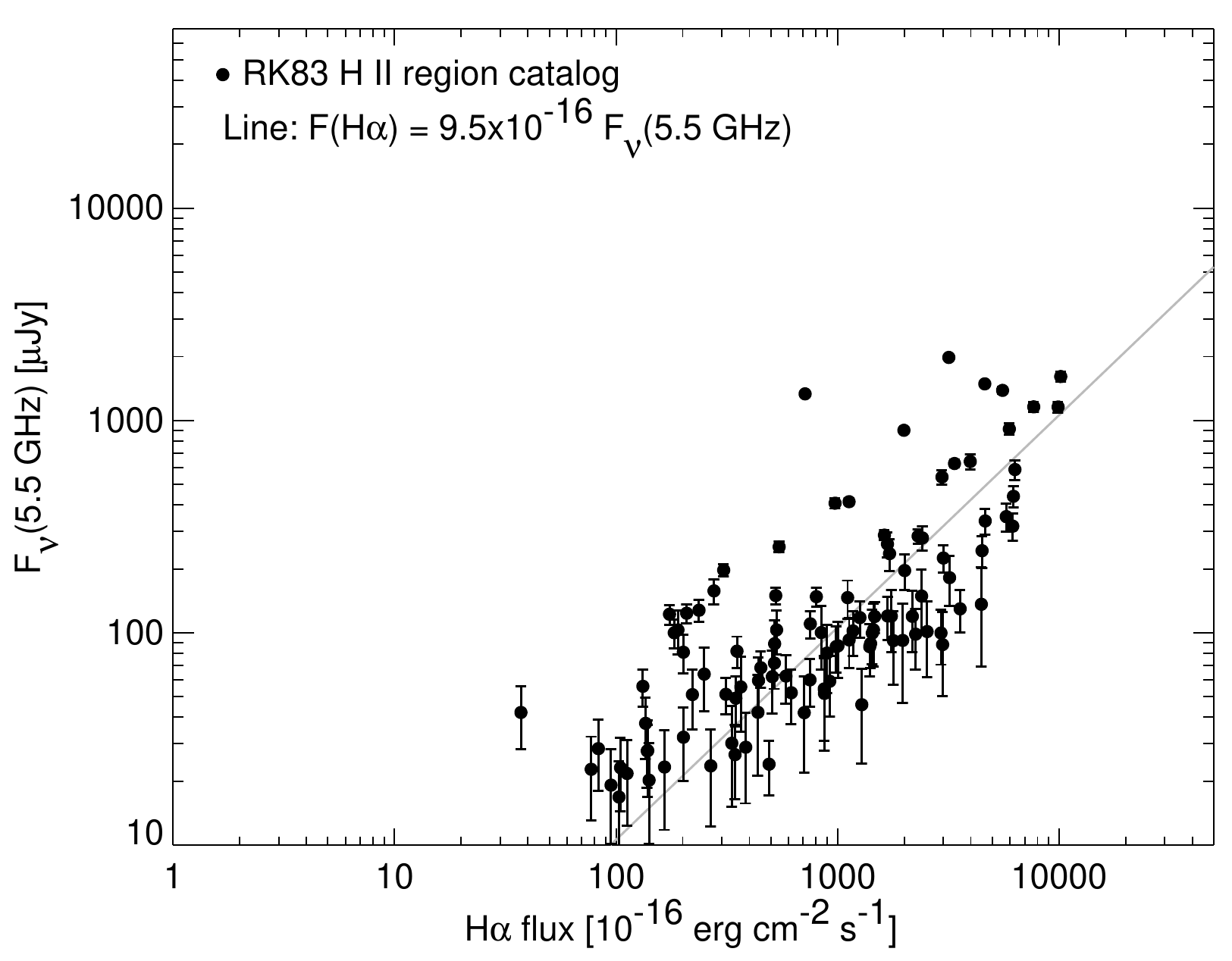}
\includegraphics[width=0.495\textwidth]{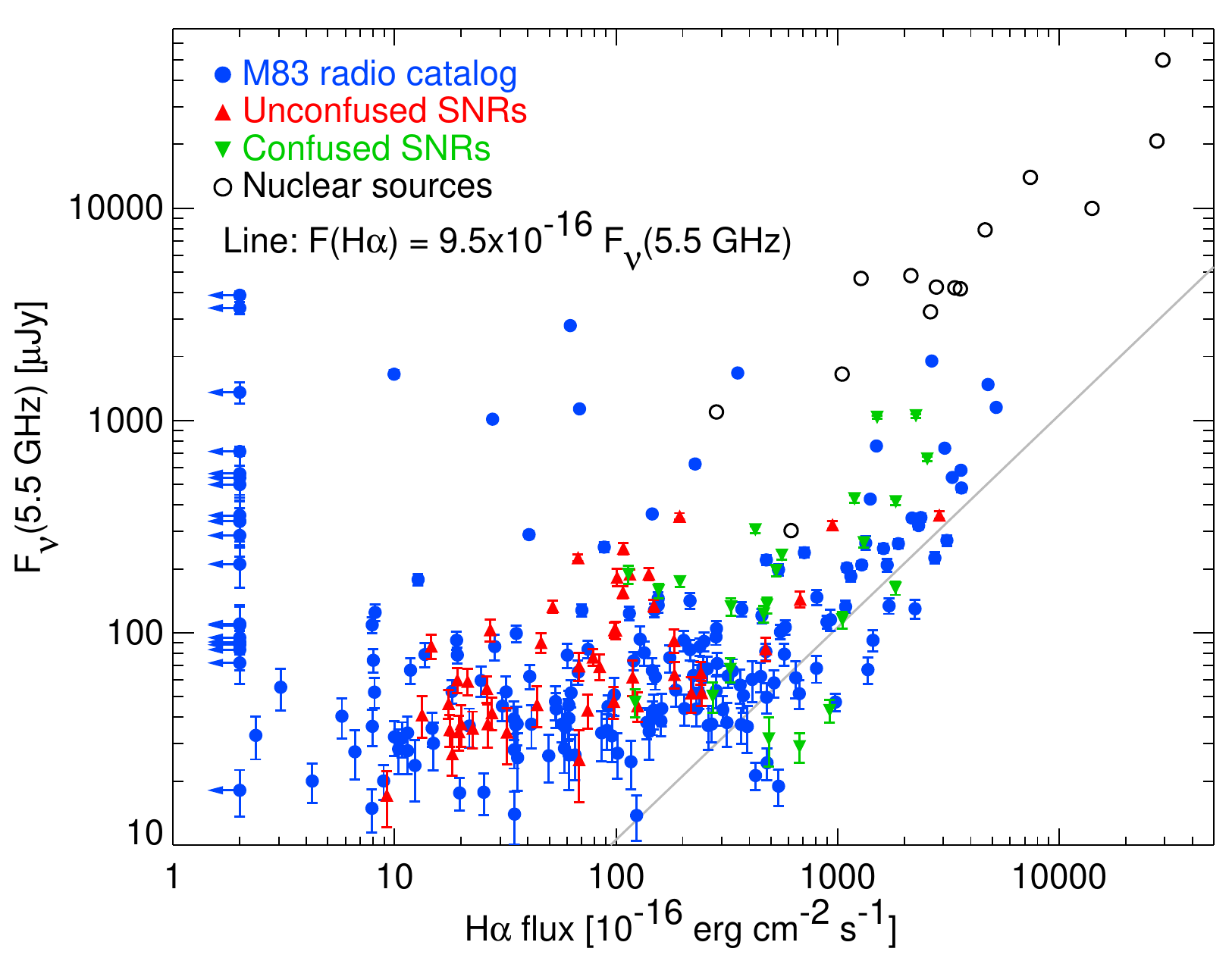}
\caption{
	Left: H$\alpha$ versus radio flux density for a catalogue of \hii\ regions taken from \citet{rumstay83}.  The line shows the expected relationship between H$\alpha$ and free-free emission at 5.5~GHz from \citet{caplan86}.
	Right: H$\alpha$ versus radio flux density for all the sources in our M83 radio catalogue. There is a population of sources distributed near the line.  SNRs in unconfused regions (red triangles) generally have excess radio emission compared with the \hii\ regions. SNRs in confused regions (green inverted triangles) often have radio emission consistent with that expected for \hii\ regions.  Open symbols are sources located within the confused nuclear region.
	}
\label{fig:ha_vs_radio}
\end{figure*}

Although the radio spectral indexes derived from our data are suspect, clearly many well-detected radio sources in the catalogue correspond with \hii\ regions, which emit via free-free emission.   The expected correlation between radio flux and \ha\ emission \citep{caplan86} provides another way to investigate the radio source population.

Figure~\ref{fig:ha_vs_radio} shows two plots of 5.5 GHz flux versus the \ha\ flux for different targets.  The radio fluxs are from our catalogue, and the \ha\ fluxes have been derived by setting corresponding regions on the Magellan \citep{blair12} continuum-subtracted \ha\ image.  The plot on the left is for a selection of \hii\ regions from \citet{rumstay83} and shows the expected correlation between these two parameters (albeit with significant scatter).  The plot on the right of the Figure shows our entire radio catalogue with several subsets highlighted.  The unconfused SNR measurements (shown in red) are shifted to the left, indicating higher radio to \ha\ ratios.  The SNRs in confused regions mainly cluster around the \hii\ region line, consistent with \hii\ emission dominating the measured radio fluxes for these objects. However, there are many other catalogue sources that appear enhanced in radio emission relative to \ha, basically overlying the SNR sources.  What are they?

\begin{figure*}
	\centering
\includegraphics[width=0.8\textwidth]{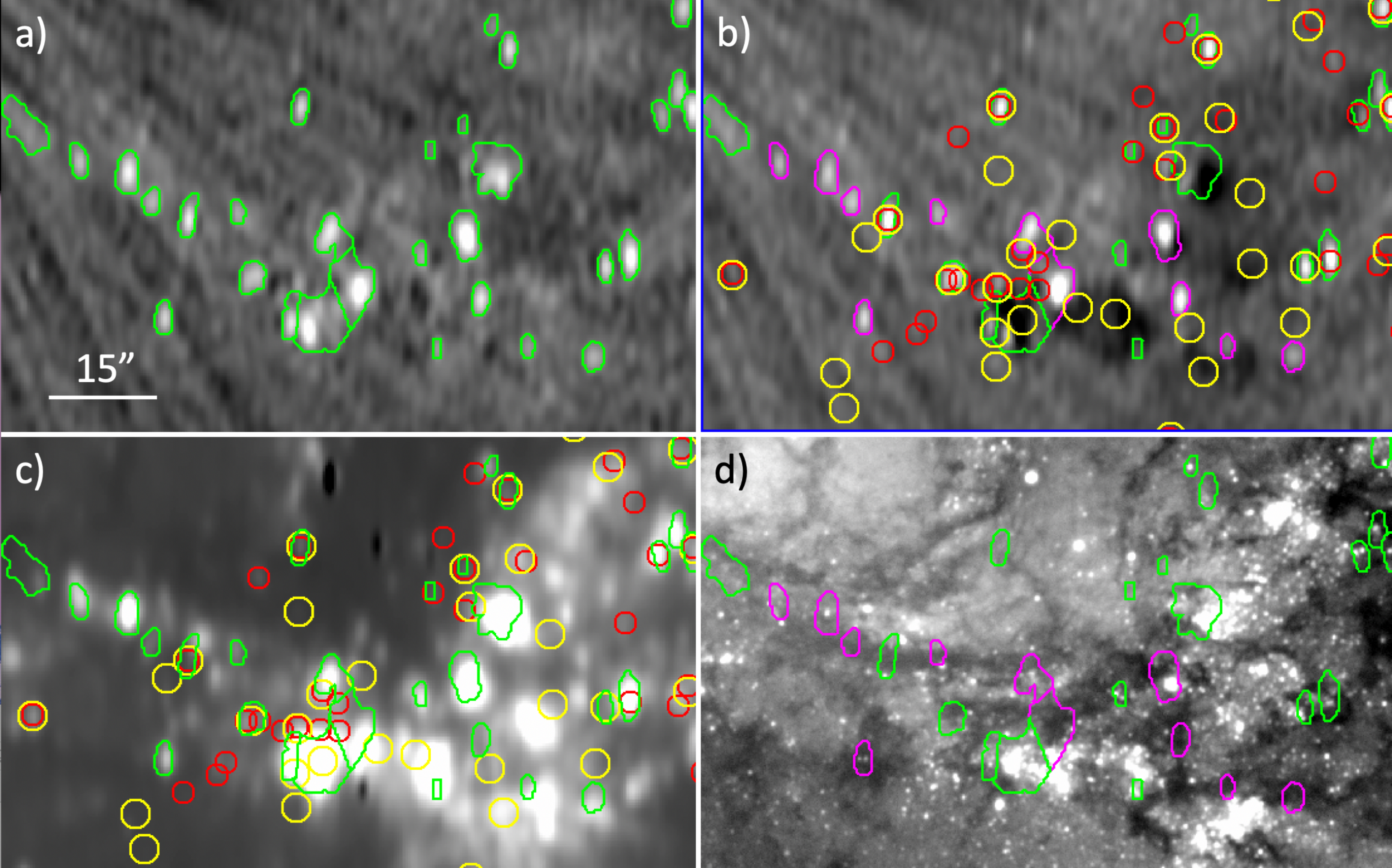}
\caption{Four-panel figure centered at RA(J2000) 13:36:54.8, Dec(J2000) -29:52:54.3, showing a region to the SW of the M83 nucleus.  For scale, the region shown is 1\arcmin\ in the N-S dimension.  Panel a shows the radio detection image with the radio islands overplotted in green.  Panel c shows the Magellan continuum-subtracted \ha\ image smoothed to the radio beam size. In  panel b, we have scaled and subtracted the \ha\ image from the radio image such that regions of primarily photoionized emission should disappear or be greatly reduced. Thus, the radio emission that remains in this panel is stronger than expected from typical photoionized emission. Some but not all of the sources that remain in panel b align with SNRs (red circles) and/or X-ray sources (yellow circles).  The ones that do not (irregular magenta regions) project almost exclusively onto dark, dusty regions, as shown in panel d, which is a visual continuum image of the region from Magellan.  Hence, these radio sources 
are either background sources or more likely radio \hii\ regions or SNRs with no detected optical or X-ray emission.  
\label{fig_subtract_ha}
}
\end{figure*}

 In Figure~\ref{fig_subtract_ha}, we show a different experiment. Panel c shows  version of the Magellan \ha\ image (appropriately matched to the radio data resolution). This was scaled and subtracted from the radio detection map in panel a, with the result shown in panel b. The white radio sources that remain in panel b thus have stronger radio emission than expected from any associated \hii\ emission.  A number of SNRs and X-ray sources align with these ``excess radio'' sources.  However, a number of them do not have counterparts.  The radio regions for these sources are shown in magenta in panels b and panel d, which is a visual band Magellan image of the region.  These magenta regions nearly all project onto dark, dusty regions.\footnote{This is true for roughly 90\% of the $\sim 75$ radio excess sources we see across the entire data set.}  Some of these sources might represent a population of radio-detected \hii\ regions or SNRs whose optical emission is simply lost due to dust.

A few of these radio excess objects lie in a portion of M83 for which WFC3 IR Pa$\beta$ observations exist.\footnote{Sadly, only two of the seven WFC3 fields observed in M83 have Pa$\beta$ data in the archive.} Pa$\beta$ is less affected by dust than \ha, and indeed evidence of \hii\ emission peeking through the dust does seem to be present for four out of the five regions we can check, making the \hii\ region explanation seem most likely for most of these sources. In principle, some of these radio excess sources could be heavily extincted SNRs, but our inability to determine radio spectral indexes does not allow us to distinguish between these explanations.

\section{Discussion}
\label{sec:discussion}

Our main motivations in undertaking this study were to obtain a better understanding of the radio properties of M83's SNRs and SNe.

\subsection{Supernova Remnants}

\begin{figure}
	\centering
\includegraphics[width=0.98\columnwidth]{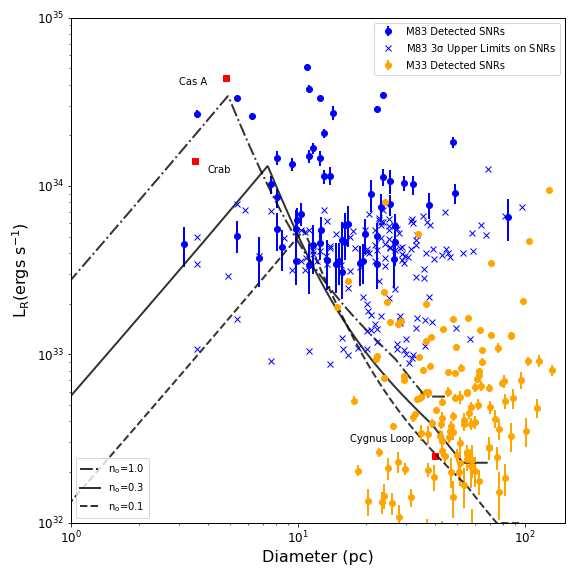}
\caption{Radio luminosities of (isolated) SNRs in M83 and M33 as a function of SNR diameter.  Upper limits for other (isolated) SNRs in M83 are also shown.   Predictions for the radio luminosity of a SNRs expanding into an ISM with a density of 1.0, 0.3 and 0.1 cm$^{-3}$ based on the models of  \citet{sarbadhicary17,sarbadhicary19} are also shown (see text).  The positions of three well-known Galactic SNRs -- Cas A. the Crab, and the Cygnus Loop -- are indicated as red squares.
\label{fig_lum_diam}}
\end{figure}

A comparison of 5.5~GHz radio luminosities as a function of SNR diameter in M83 and M33  is shown in Figure~\ref{fig_lum_diam}.    Here we have used the forced photometry results and included only the M83 SNRs that were located in relatively unconfused portions of the radio images (excluding the SNRs with a C classification). For context, the positions of three well-known Galactic SNRs -- Cas A, the Crab, and the Cygnus Loop -- are also shown.   The brighter SNRs observed in M83 are as bright as the brightest, young SNRs we observe in the Galaxy.  All of the objects we detect in M83 are considerably brighter than older Galactic SNRs like the Cygnus Loop, which would have fallen below our detection limit.

The distribution of radio luminosities at any particular diameter is quite broad, at least an order of magnitude, even if one disregards the upper limits.  The detected SNRs in M83 are brighter than those in M33.  This is partly due to the differing sensitivities of the VLA survey of M33 and the ATCA survey of M83, and also the different distances to the two galaxies.  However, that is not the whole story, since remnants as bright as those detected in M83 would certainly have been detected in the M33 survey if they were present. It is more likely that some of the difference between the two galaxies arises from the fact that there are many more small diameter (hence, young) SNRs in M83 than there are in M33, and SNRs with smaller diameters tend to be brighter.  Indeed, the median diameter of SNRs detected at radio wavelengths in M83 is only 14 pc, whereas the median diameter of those that were not detected is 24 pc.  For comparison, the median diameter of the SNRs detected at radio wavelengths in M33 is 51 pc.  It is also true that the diameter distribution of the entire sample of SNRs in M83 is systematically smaller than in M33.  \citet{winkler17} have argued that this difference and the plethora of small diameter SNRs in M83 arises from differences in the properties of the ISM, most notably the typical density,  in M33 and M83.  SNRs expanding into higher density gas evolve more rapidly to the Sedov and then radiative phases. Since the total energy radiated remains about the same, such SNRs have peak luminosities that are higher than those expanding into more tenuous gas.  

The apparent trend  of SNR luminosity  with diameter (with much scatter) in Figure~\ref{fig_lum_diam} is fairly consistent with models for the radio luminosity, such as those of  \citet{sarbadhicary17,sarbadhicary19}. These models are based on standard  diffuse shock acceleration (DSA), which creates a power-law distribution of electrons with a power-law slope $p$ of about 2.2 and an acceleration efficiency that is a fraction $\epsilon_e$  of the post-shock energy density ($\sim \rho v_s^2$, where $v_s$ is the shock velocity).  As a result, the number of relativistic electrons available is a strong function of the shock velocity.  They assume that the interstellar magnetic field varies as $\rho^{0.47}$  and that the magnetic field is amplified behind the shock to a value that is roughly proportional to the post-shock energy density.  These assumptions allow one to calculate the synchrotron spectrum of a SNR, assuming dynamics from one dimensional models.  The models are fully specified by the energy of the SN explosion, the SN type, the ejecta mass, the density of the ISM, the electron acceleration efficiency, the electron spectral index, and the interstellar magnetic field at a reference density of 1 $\rm cm^{-3}$.  \citet{sarbadhicary17} found that they could account for the radio luminosity function for M33, as it existed in \citet{gordon99}, using a model with a SN rate of 0.3 per century and an electron acceleration efficiency of 0.0042. 

Three such models are shown in Figure~\ref{fig_lum_diam}.\footnote{For the calculations here we have used revised versions of the model, which take account errors reported in \citet{sarbadhicary19}, calculated with routines provided by S. K. Sarbadhicary in the Github repository located at \url{https://github.com/sks67/s17lc}.}  All of the models are for a core-collapse SN with an explosion energy of \EXPU{1}{51}{ergs} and ejecta mass of 2\MSOL, an electron spectral index of 2.2, and an electron acceleration efficiency  $\epsilon_e$  of 0.0042.   Lower values of the efficiency yield somewhat lower luminosities. Higher luminosities would be predicted for higher explosion energies. Interestingly all of the models predict about the same radio flux  between diameters of 10 and 20 pc, so if the models are correct, the large variation in radio luminosity at a particular diameter is not primarily a density effect.  Changing the ejected mass also does not move the curves toward the upper right because once the Sedov phase is reached, the swept-up ISM mass dominates the evolution and the ejected mass becomes irrelevant. We note that the majority of the SNRs both in M83 and in M33 lie above the model curves (i.e. at higher radio luminosities).  This may  in part be due to the sensitivity of the existing observations, but if that is the case, it only exacerbates the problem of explaining the large spread of luminosities at any particular diameter.  

\begin{figure}
	\centering
\includegraphics[width=0.495\textwidth]{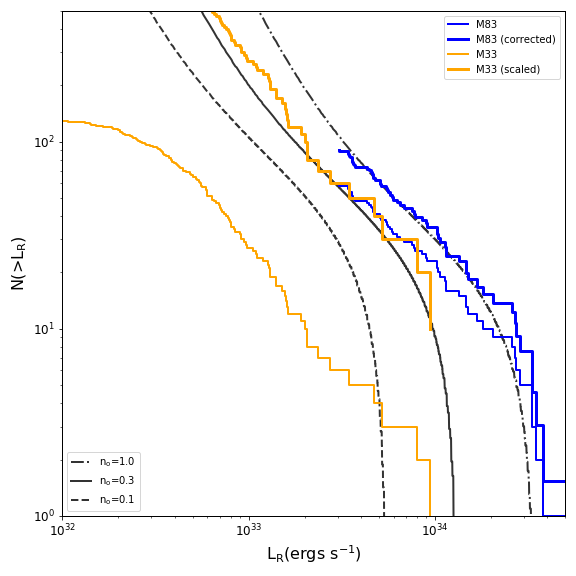}
\caption{
A comparison between the radio luminosity function of SNRs detected in  M83  and M33 at 5.5 GHz.  The M33 measurements were made at 1.4 GHz and have been adjusted to 5.5 GHz assuming a radio spectral index in $f_\nu$ of -0.6.  In the case of  M83, the thinner line represents the luminosity function of isolated SNRs and the thicker line is corrected for the fraction of SNRs that were in complex regions of the radio image.  For M33, the thinner line is the observed luminosity function and the thicker line  is the luminosity function scaled to the SFR of M83.  The three (black) lines are the predicted luminosity functions for the same three models presented in Figure~\ref{fig_lum_diam}.
\label{fig_lumfunc}}
\end{figure}

A comparison of the luminosity function of SNRs in M83 to those in M33 is shown in Figure~\ref{fig_lumfunc}. Both luminosity functions are roughly power laws.  \citet{chomiuk09} have argued that the luminosity function of SNRs is generally a power law, and that seems to be the case in both of these galaxies.  The normalization of the luminosity function should be determined by the star formation rate (SFR), which after all determines the SN rate in a galaxy. Estimates of SFR vary for M33\footnote{\citet{williams18} provides three different SFR estimates for M33: $0.17\pm0.06$\MSOL\  yr$^{-1} $ from multi-wavelength analysis, to 0.25$^{+0.10 }_{-0..07}$\MSOL\  yr$^{-1} $ from FUV and 24 $\mu$m imaging, to 0.33$^{+0.05}_{-0.06}$\MSOL\  yr$^{-1} $ using SED fitting. \citet{verley09} estimate  $0.45\pm0.10$\MSOL\  yr$^{-1} $ using a combination of UV, optical, and IR data.} but are approximately 10$\times$ less than for M83.  In Figure~\ref{fig_lumfunc}, we show curves where the M33 curve has been adjusted to the M83 SFR, and the M83 curve has been corrected for the fact that some SNRs in M83 are in radio-confused regions. With these adjustments, the luminosity functions for the two galaxies are actually fairly similar.  
 
 \begin{table*}
\caption{Radio flux densities of the M83 historical SNe SN1923A, SN1945B, SN1950B, SN1957D, SN1983N, and B12-17a. All reported flux densities are in units of \mJybeam. For completeness, we show the 6cm 1984--1998 VLA flux densities from \citet{maddox06}. All other reported flux densities were measured at 5.5~GHz from our ATCA radio images. We note that the seventh SN1968L is located within the bright nuclear region, meaning that no strong constraints could be placed from our data.
.}
\centering
\label{tab:historical_SNe}
\begin{tabular}{ccccccc}
\hline
Epoch &  SN1923A$^a$ & SN1945B & SN1950B & SN1957D & SN1983N & B12-174a\\ 

\hline
1984--1985 & --- & --- & 0.43 $\pm$ 0.06 & 2.18 $\pm$ 0.06 & 1.30 $\pm$ 0.12 & 0.75 $\pm$ 0.09 \\

1990 & 0.28 $\pm$ 0.04 & --- & 0.54 $\pm$ 0.05 & 1.57 $\pm$ 0.04 & --- & 0.42 $\pm$ 0.05 \\
1998 & 0.19 $\pm$ 0.05 & --- & 0.50 $\pm$ 0.04  & 0.60 $\pm$ 0.05 & --- & 0.31 $\pm$ 0.05 \\
2011 & 0.115 $\pm$ 0.02 & $<$0.075 & 0.32 $\pm$ 0.02 &  0.30 $\pm$ 0.03 & --- & 0.28 $\pm$ 0.02 \\
2015 & 0.078 $\pm$ 0.008 & $<$0.025 & 0.31 $\pm$ 0.01 & 0.195 $\pm$ 0.01 & --- & 0.26 $\pm$ 0.01 \\
2017 & 0.105 $\pm$ 0.015 & $<$0.060 & 0.33 $\pm$ 0.025 & 0.160 $\pm$ 0.02 & --- & 0.23 $\pm$ 0.02 \\

\hline
\multicolumn{7}{l}{  
\begin{minipage}{0.9\textwidth} 
$^a$ the SN1923A position has been reconstructed based on reference to the original publication, and comparison to modern data and coordinates (accurate to a few arcseconds). Here, the listed information corresponds to A096 from L14, which is $\sim1.6$\arcsec\ east of this reconstructed position.
\end{minipage}
}
\end{tabular}
\end{table*}

\begin{figure*}
\centering
\includegraphics[width=0.7\textwidth]{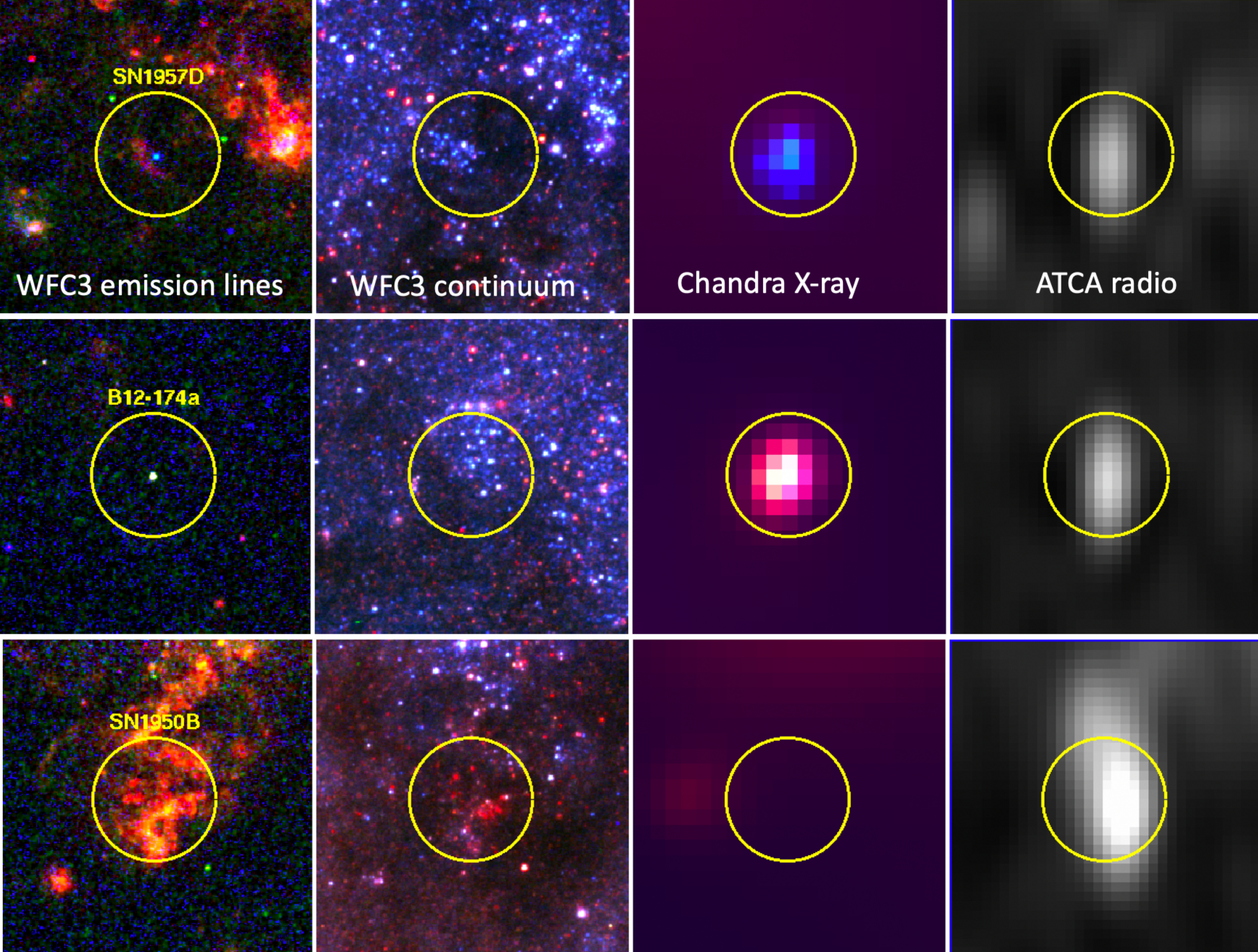}
\caption{Three 4-panel figures showing multiwavelength imagery for the three historical M83 SNe that show radio emission, SN1957D (top), B12-174a (middle) and SN1950B (bottom).  The four panels show (from left to right) {\it HST} WFC3 subtracted emission lines (\ha\ red, \sii\ green, \oiii\ blue); {\it HST} WFC3 continuum (I band red, V band green, B band blue), {\em Chandra} X-ray (soft red, medium green, hard blue), and the ATCA detection image.  The yellow circles are 4\arcsec\ in diameter.
\label{fig_hist_sne}}
\end{figure*}
 
The predicted luminosity functions based on the models of \citet{sarbadhicary17,sarbadhicary19} are also shown in Figure~\ref{fig_lumfunc}, assuming a SN rate of six per century for M83.  Given the uncertainties, it is interesting how well the predicted and observed luminosity  functions agree.  That is somewhat surprising given that many of the detected SNRs in Figure~\ref{fig_lum_diam} have luminosities well above those predicted by the model. It is an indication, in part, that luminosity functions do not have the information content required to fully validate detailed models for radio emission from SNRs. That noted, one of the more interesting features of the comparisons is that both the model and the data (for M83)  appear to show a luminosity cutoff.  The luminosity cutoff in the model is quite sensitive to $\epsilon_e$ as well as to the explosion energy.  Whether the cutoff is significant is difficult to tell, however, because SNRs reach their highest luminosity in the model at the transition between the free expansion and Sedov phases of their evolution, where the theory is likely to be most uncertain.

\subsection{Historical Supernovae}
\label{sec:histsn}

Studies of radio SNe can provide important constraints on models describing the time evolution of these objects \citep[e.g.,][]{cowan85,stockdale01,stockdale06}. The rise and decay rates, and the time after explosion to the peak of the radio emission provides information on the circumstellar mass-loss rate for the progenitor, helping to reveal the late stages of stellar evolution \citep[e.g.,][]{chevalier1984}. Typical models of the radio emission from such SNe predict an initial brightening of the radio emission, peaking after tens to hundreds of days, after which a general decline in the radio luminosity is observed, thought to be due to a declining density of the CSM  \citep[e.g.,][]{chevalier1984,weiler02}. Following a $\gtrsim100$-year fading period, the radio emission may then re-brighten as it collides with the surrounding ISM and evolves into a SNR, before flattening over time or slowly fading \citep{cowsik84}.  The models of \citet{sarbadhicary17,sarbadhicary19} assume a constant density for the ISM, and would only apply once interactions with the local CSM have ceased to be important.

\begin{figure}
\centering
\includegraphics[width=1\columnwidth]{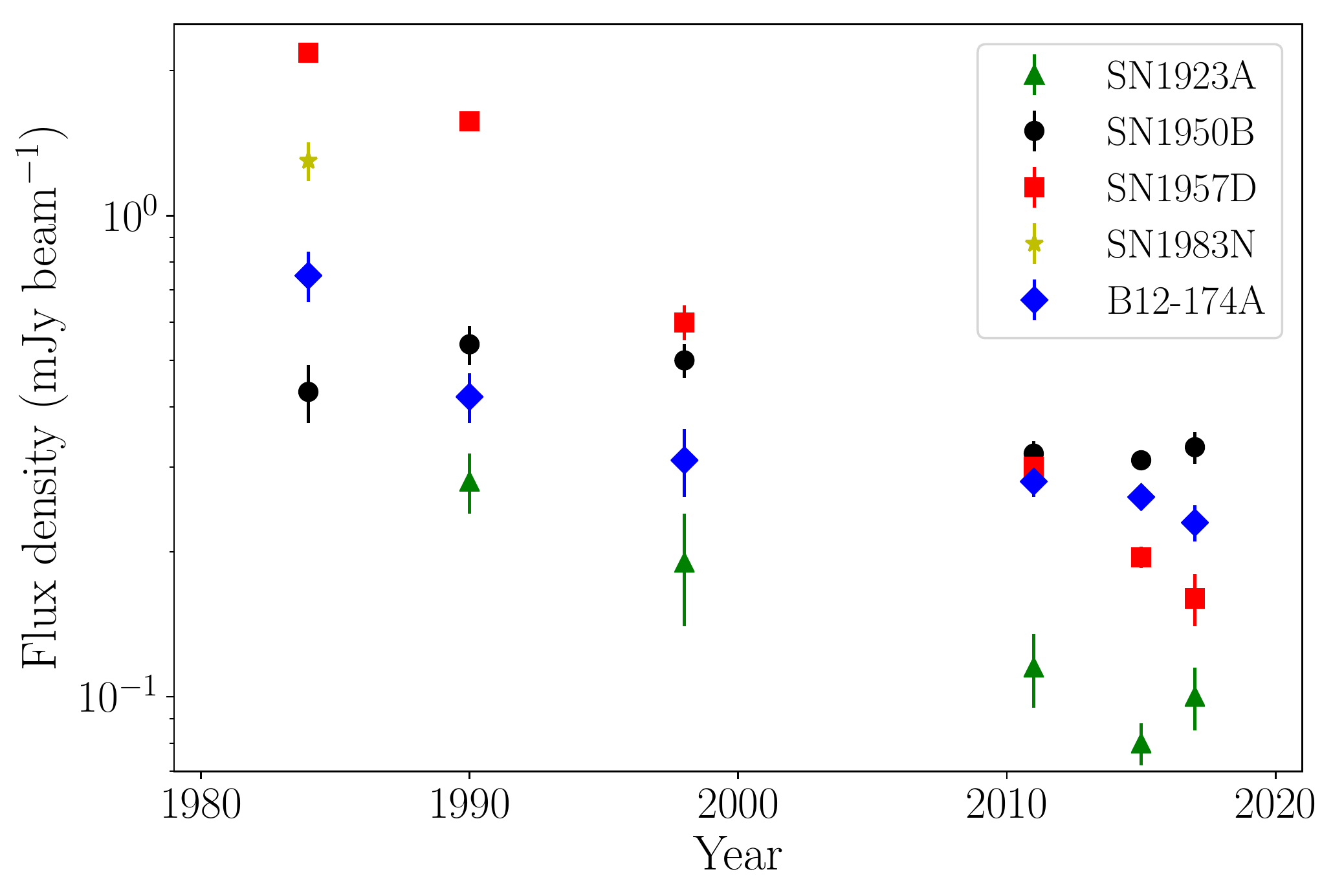}
\caption{The radio light curves of the five historical SNe detected by \citet{maddox06} and/or in our radio images. To match the frequency of the peak 6\,cm flux densities reported in \citet{maddox06} as closely as possible, we report the peak 5.5~GHz flux density from our radio observations taken in 2011, 2015, and 2017. In all cases but SN1950B the radio emission is seen to be fading steadily over time. For SN1950B, while the radio emission is generally decreasing, two epochs appear brighter than the previous, possibly due to the differing observational setups and varying amounts of \hii\ region contamination being sampled.}
\label{fig_hist_sne_time}
\end{figure}

Six SNe have been observed in M83 since 1923. As part of our earlier work identifying SNRs, we also identified one additional object, B12-174a, whose characteristics are consistent with it also being a young SN less that 100 years post-explosion \citep{blair15}. Of these seven intermediate-age SNe, three are clearly detected in our radio images (see Table~\ref{tab:historical_SNe} and Figure~\ref{fig_hist_sne}) and possibly a fourth (SN1923A -- see below), while a fifth was detected in an earlier epoch but has since faded below our detection threshold (SN 1983N). Here, we report the time-evolution of the radio emission from these historical SNe (Figure~\ref{fig_hist_sne_time}) and comment on the additional SNe that were not detected.

SN1923A is in a confused \hii\ region with bright stellar sources. The position of this SN is not well known.  The original report from C. O. Lampland in 1923 only provides an offset from the nucleus, but the nucleus is ill-defined.  However, \citet{pennington82} reproduce a copy of the SN discovery plate in their Figure 1b; we have reconstructed the position by comparing this image to our earlier imaging data with a good astrometric solution \citep{blair12}. Even so, the derived position is  thought to be accurate only to a few arcseconds. No obvious optical counterpart is seen near this position, nor is there an X-ray counterpart. In the radio, we detect two sources within $\sim1.6$\arcsec\ of the reconstructed coordinate, both of which appear to be fading over time. Although it is possible that both radio lobes are somehow related to the SN, here we report the results from the brighter radio counterpart that spatially matches that of the source A096 from \citet{long14}. Since first reported by \citet{maddox06}, this source has been fading steadily over time, from 0.28\,\mJybeam\ in 1990 to $\sim0.10$~\mJybeam\ in our 2017 data.

SN 1945B is located in the outskirts of the galaxy and has little stellar or other emission nearby. No X-ray counterpart is detected and no radio emission is detected within our images that could be associated with this source.

The nominal position for SN 1950B places it within an \hii\ region (see Figure~\ref{fig_hist_sne}). While no optical remnant from the SN has been recovered, we detect a bright radio counterpart to SN 1950B.  The radio emission from this source has generally faded over time, though source confusion (between diffuse emission associated with the \hii\ region and the interaction of the SN shock with its surroundings) limits the quantitative assessment of its decline.

SN1957D (see Figure~\ref{fig_hist_sne}) has an optical counterpart that was discovered in late 1980's that has been fading over time \citep[][and references therein]{long12}. It is an unresolved \oiii\ point source to {\it HST} and thus has an excellent position. 
This SN is associated with a faint, hard X-ray source, which may be a pulsar wind nebula \citep[PWN;][]{long12}. SN1957D is also a bright radio emitter; \citet{maddox06} reported \mJybeam\ flux densities that were fading over their 14 year observation span. With our ATCA observations, we show that the source has continued to fade steadily over time (Figure~\ref{fig_hist_sne_time}). 

SN 1983N occurred within a cluster of young stars;  no known optical or X-ray counterpart has been detected.  \citet{maddox06} reported a detection in radio observations taken in 1984/85 but not in their subsequent radio observations in 1990 and 1998. We did not detect it in our ATCA radio observations. 

We detect bright radio emission from the recently discovered SN B12-174a (see Figure~\ref{fig_hist_sne}). Looking back at the previous VLA data \citep{maddox06}, the source was clearly detected, and our ATCA observations show that the 6cm/5.5~GHz radio emission has faded since 1984 (Figure~\ref{fig_hist_sne_time}). {\it HST} optical observations \citep{blair15}  show this source has a point-like counterpart. Unlike the hard X-ray source seen in SN1957D, there is a soft X-ray source coincident with B12-174a.

The remaining one of these seven recent SNe, SN 1968L, is located in the radio-bright, complex nuclear region.  The precise position of the SN is not known,\footnote{\citet{dopita10} found a possible optical counterpart in {\it HST} data, but given the complexity of the region and the difficulty in subtracting the continuum, this marginal source detection is far from solid.} therefore, we are unable to attach meaningful constraints for its radio emission in the confused nuclear region of M83.

Our radio results agree well with typical models for radio emission from SNe, which predict steadily decreasing radio emission \citep[e.g,][]{weiler02}.   We also searched for any significant radio brightness variations from all SNR listed in our radio catalogue. Results show that the population of the detected radio SNRs appeared to be steady over the 6-years spanning our ATCA observations, typically remaining within $\sim$2$\sigma$ between ATCA epochs. 

\section{Summary and Conclusions}

\label{sec:summary}

We have presented a new, deep radio survey of M83 using three epochs of ATCA data at 5.5 and 9 GHz.  From these data, we have produced a catalogue of radio sources in the galaxy M83 containing 270 sources, all but four of which were detected at 4$\sigma$ or greater. With a detection limit of 14 $\mu$Jy, this is by far the most sensitive radio survey of this important face-on spiral galaxy. 

The sources are concentrated along the spiral arms of the galaxy and mostly arise from emission from \hii\ regions and SNRs, or combinations thereof.  The contribution from background sources is small.  Even with the spatial resolution provided by ATCA, the radio emission from the starburst nuclear region is not totally resolved.

In addition to the general radio source catalogue, we have estimated the radio flux directly at the positions of known optical SNRs and known X-ray sources in M83. By visually inspecting the radio images, we have separated the sources into those for which the radio flux density is a meaningful measurement (or upper limit) of the radio flux density for the source (203 of 304 SNRs and 350 of 424 X-ray sources), and those for which the radio measurement is most likely dominated by continuum emission from associated \hii\ regions (101  SNRs and 74 X-ray sources).  We detect radio emission at the position of 64  SNRs and 75 X-ray sources.  A large majority of the X-ray sources that are detected as radio sources are actually SNRs.

We attempted to locate candidate radio SNRs by searching for radio sources with excess emission over what was expected from local \ha\ emission. While we did find some objects with radio excess, the majority of these objects lie along lines of sight with substantial visible absorption and are most likely \hii\ regions behind dust, not candidate radio SNRs.

As was the case in M33 \citep{white19}, there is a large variation in radio luminosity at any specific SNR diameter.  The SNRs we detect in M83 are generally brighter and smaller in diameter than those in M33.  The radio luminosity function is a power law as has been reported for other galaxies, and the normalization of the luminosity function in M33 and M83  scales roughly as the relative star formation rates. 

Of the six historical SNe in M83, two (SN1950B, and SN1957D) are clearly visible in our radio images and a third (SN1923A) is probably seen.  All three are declining in brightness with time.  Additionally, we have detected radio emission from the very young SNR B12-174a, which is likely the result of an unobserved  SN in the last century \citep{blair15}.  This source was detected in the earlier VLA data as well, although not recognized previously, and it also appears to be clearly fading with time.  

Although we have observations at both 5.5 GHz and 9 GHz, the derived spectral indices  are not reliable, due primarily we believe to differences of coverage in the UV plane at the two frequencies and structure being partially resolved out at the higher frequency.  Not only were the spectral indices uniformly steeper than expected, but we do not see the sort of bimodal distribution of spectral indices seen in M33 \citep{white19}.   This made it impossible to identify potential SNRs in M83 based on their radio properties alone. Our inability to resolve this issue provides a cautionary lesson in the interpretation of spectral indices for other studies of radio emission in nearby galaxies.  Unless one can cleanly separate thermal and non-thermal sources, one cannot expect to identify SNRs in external galaxies via radio properties alone.

\section{Acknowledgements} 
We thank the anonymous referee for their helpful comments and suggestions. TDR also thanks Leith Godfrey, James Miller-Jones, and Paul Hancock for useful discussions pertaining to the radio analysis. TDR acknowledges support from the Netherlands Organisation for Scientific Research (NWO) Veni Fellowship, grant number 639.041.646. RLW acknowledges research support from NASA through the Space Telescope Science Institute. KSL, WPB, and PFW acknowledge that partial support for the analysis of the data was provided by NASA through grants No.~HST-GO-14638 and No.~HST-GO-15216 from the Space Telescope Science Institute, which is operated by AURA, Inc., under NASA contract NAS 5-26555.  WPB acknowledges additional support from the Johns Hopkins University Center for Astrophysical Sciences and the dean of the Krieger School of Arts and Sciences during this work. PFW acknowledges additional support from the National Science Foundation through grant AST-1714281. The Australia Telescope Compact Array (ATCA) is part of the Australia Telescope, which is funded by the Commonwealth of Australia for operation as a National Facility managed by CSIRO. This research has made use of NASA's Astrophysics Data System and the Mikulski Archive for Space Telescopes.

\bibliographystyle{mnras}
\bibliography{m83}

\clearpage
\newpage

\begin{center}
\appendix
\section{The M83 Nucleus}
\label{sec:nuc_appendix}
\end{center}

In this Appendix, we provide a more detailed discussion on the rich literature pertaining to the starburst nuclear region.  The main unsolved problems for the nuclear structure of M83 are a) whether this galaxy has a supermassive BH, and b) if so, where it is located.  In the absence of a smoking gun telling us where the BH is located, we need to use scaling relations to estimate a plausible range of nuclear BH mass that might be expected for this galaxy.

There are a number of ways to parlay observed parameters in the nuclear region or galaxy into estimates of or limits on any massive BH present.  Below we outline several of those estimates, which unfortunately vary between methods; this is quite characteristic of late-type spiral galaxies.  The bottom line is, we can say that M83 is expected to have a small nuclear BH, probably with a mass near $\sim10^6 M_{\odot}$.  The details of the specific estimates follow.

\smallskip
\smallskip

\subsection{Mass of the Nuclear BH from Scaling Relations}

A stellar velocity dispersion $\sigma \approx 75$ km\,s$^{-1}$ in the nuclear region \citep{houghton08} corresponds to a BH mass $M_{\rm {BH}} \approx 4.5 \times 10^5 M_{\odot}$, from the $M_{\rm{BH}}$-$\sigma$ scaling relation of \citet{davis17}. The relation has a total dispersion of 0.63 dex. A similar velocity dispersion $\sigma = (78 \pm 10)$ km\,s$^{-1}$ was measured by \citet{devaucouleurs83}, corresponding to $M_{\rm {BH}} \sim 5.6 \times 10^5 M_{\odot}$, with an uncertainty of a factor of 2.

From a {\sc galfit} decomposition of the infrared emission into disk, bar and bulge components, \citet{piqueras12} obtain a $K$-band bulge luminosity $\log \left(L_{K,{\rm {bulge}}}/L_{K,\odot} \right) \approx 9.71$. Converting this value to a stellar mass with the  relation $M/L_K = 0.62$ \citep{davis19a,graham19} we obtain $M_{\ast,\rm{bulge}} \approx 3.1 \times 10^9 M_{\odot}$ and $M_{\rm {BH}} \approx 7.3 \times 10^5 M_{\odot}$ (from the $M_{\ast,\rm{bulge}}$-$M_{\rm {BH}}$ scaling relation of \citealt{davis19a}, with an intrinsic root-mean-square scatter of 0.70 dex).

For late-type spiral galaxies where the definition of a spheroidal bulge is somewhat uncertain, an alternative scaling relation between $M_{\rm{BH}}$ and total stellar mass $M_{\ast,\rm{tot}}$ was proposed by \citet{davis18}. Using the integrated $K$-band luminosity $L_K \approx 6.1 \times 10^{10} L_{K,\odot}$ \citep{jarrett03}, we estimate $M_{\ast,\rm{tot}} \approx 3.8 \times 10^{10} M_{\odot}$, and this corresponds to $M_{\rm {BH}} \approx 4.5 \times 10^6 M_{\odot}$ (with a scatter of 0.66 dex).

The maximum rotational velocity of the disk provides another scaling relation with $M_{\rm {BH}}$ in spiral galaxies, with a small scatter of 0.58 dex \citep{davis19b}. For M83, \citet{lundgren04} measured $v_{\rm {rot,max}} \approx 190$ km s$^{-1}$, which corresponds to a predicted $M_{\rm {BH}} \approx 5.8 \times 10^6 M_{\odot}$.

Again for spiral galaxies, a relation between $M_{\rm {BH}}$ and pitch angle of the spiral arms was proposed by \citet{davis17}. For M83, the observed pitch angle is $\phi = 11^{\circ}.8 \pm 2^{\circ}.7$ (B.~L.~Davis, priv.~comm.), which corresponds to $\log(M_{\rm{BH}}/M_{\odot}) = 7.57 \pm 0.58$ (including the scatter of 0.43 dex of this relation). This estimate is clearly an outlier compared with the other scaling relations.

\subsection{Location of the Nuclear BH}

There are two particular locations inside the starburst nuclear region that have been discussed as the candidate nucleus in the literature \citep{knapen10,muraoka09,houghton08,diaz06,sakamoto04,thatte00}: i) the ``optical nucleus''; and ii) the photometric and kinematic nucleus.  The optical nucleus is a bright, heavily reddened star cluster (see L14-233 in Figure~\ref{fig_nucleus_6pan}). Its dynamical mass is $M_{\rm {cl}} = (1.1 \pm 0.4) \times 10^7 M_{\odot}$ \citep{piqueras12,thatte00}. From its photometric and spectroscopic properties, \citet{piqueras12} also derived an alternative mass value $M_{\rm cl} = (7.8 \pm 2.4) \times 10^6 M_{\odot}$ (consistent with the estimates of \citealt{wiklind04} and \citealt{thatte00}), with an age of $\sim 10^8$ yrs. Its intermediate age (an order of magnitude higher than the other star clusters in the starburst nucleus see, cf. \citet{wofford11}) and exceptionally high mass make this object an obvious candidate for a nuclear star cluster if not the location of a purported BH. If we apply the scaling relations between nuclear star cluster masses and various properties of their host galaxies \citep{scott13,graham12}, we notice that a nuclear cluster mass $\sim 10^7 M_{\odot}$ is indeed typical of galaxies with similar properties to M83 and nuclear BHs of $\sim 10^6 M_{\odot}$ (see also \citealt{neumayer20}).

On the other hand, the photometric center, derived from elliptical fits to the surface brightness isophotes between 12\arcsec\ and 30\arcsec, is located $\sim 3\farcs1$ to the west and $\sim 1\farcs8$ to the south of the optical nucleus \citep{knapen10}. The kinematic center, derived from models of the rotational motion on kpc scales, is also consistent with the photometric center \citep{knapen10,fathi08,sakamoto04}. There is no X-ray, optical, or infrared source at this location, and no enhancement in the velocity dispersion. Thus, the presence of a nuclear BH at that location has no direct observational support and seems unlikely.

Our X-ray and radio studies of the nuclear region can help to constrain the properties of any nuclear BH. The candidate nuclear star cluster contains the most luminous point-like X-ray source in the region (L14-233 from \citet{long14}), with a power-law-like, featureless spectrum, consistent with an accreting source rather than an SNR; it has a photon index $\Gamma \sim 2$ and a 2--10 keV luminosity of $\sim 3\times 10^{38}$ erg s$^{-1}$ \citep{yukita16,long14,soria03}. 
Instead, there is no X-ray source at the location of the photometric/kinematic nucleus. From the combined 790 ks dataset of \citet{long14}, we estimate a 2--10 keV upper limit $L_{2-10} \lesssim 1 \times 10^{36}$ erg s$^{-1}$ (corresponding to an Eddington ratio $\lesssim 10^{-8}$ for a $\sim 10^6 M_{\odot}$ BH). If the nuclear BH is located there, such low X-ray luminosity is surprising, considering the amount of gas that is in principle available in the region from large-scale inflows. The unlikely possibility that there is a hidden active nucleus completely obscured in the {\it Chandra} band ($N_H>10^{24}$ cm$^{-2}$) is ruled out by the {\it NuSTAR} study of \cite{yukita16}, who did not find any evidence for it, and determined an upper limit $L_{10-30{\rm {keV}}} \approx 10^{38}$ erg s$^{-1}$ from the unresolved nuclear region.

In the radio bands, there is a faint enhancement at the position of the optical nucleus (Figure~\ref{fig_nucleus_6pan}), visible in the 9 GHz map shown, although the spatial resolution is not sufficient to determine whether it is due a single source or diffuse emission. Regardless, we can say that the 9 GHz flux density from the optical nucleus is $\lesssim 0.2$\,mJy above the local background. The photometric/kinematic nucleus is located at the southern end of an arc of enhanced radio emission (inner edge of the starburst ring). We estimate a 9 GHz flux density of $\sim 0.3$ mJy above the local background, but there is no evidence of a compact source at that location.

Assuming a flat spectrum (typical of compact galactic nuclei), we can take 0.2 mJy also as the estimate or upper limit for the 5 GHz emission from the optical nucleus. We can then apply the fundamental plane \citep[e.g.,][]{2004A&A...414..895F,plotkin12} relation described by \citet{gultekin19}, where
\begin{eqnarray*}
&&\log \left(M_{\rm{BH}}/10^8\,M_{\odot}\right)  
 = 0.55 \pm 0.22 \\
& + &(1.09 \pm 0.10) \, \log \left(L_{5{\rm{GHz}}}/10^{38}\,{\rm {erg~s}}^{-1}\right) \\
& -& (0.59 \pm 0.16) \, \log \left(L_{2-10{\rm{keV}}}/10^{40}\,{\rm {erg~s}}^{-1}\right). 
\end{eqnarray*}
We conclude that if the observed radio and X-ray luminosities of L14-233 are from a nuclear BH, its mass must be $\lesssim 10^6 M_{\odot}$ (formally, $M_{\rm {BH}} \lesssim 3 \times 10^5 M_{\odot}$, but with a 1-$\sigma$ intrinsic mass scatter of 1 dex; \citealt{gultekin19}). On the other hand, the lack of an X-ray detection and the uncertainty in the compact radio luminosity at the photometric/kinematic nucleus mean that we cannot place any constraints on the putative BH mass at that location.

\clearpage
\newpage

\end{document}